\documentclass[twocolumn,secnumarabic,amssymb, nobibnotes,nofootinbib, aps, prd]{revtex4-1}
\pdfoutput=1

\usepackage{graphicx}
\usepackage{enumitem}
\usepackage{latexsym}
\usepackage{amsfonts}
\usepackage{amssymb}
\usepackage{color}
\usepackage{amsmath}
\usepackage{slashed}
\usepackage{dcolumn}
\usepackage{verbatim}

\definecolor{darkblue}{rgb}{0,0,0.5}
\usepackage[
pdfauthor={Nirmal Raj}]{hyperref}

\DeclareGraphicsExtensions{.pdf,.png,.jpeg}

\definecolor{orange}{rgb}{1,0.5,0}
\definecolor{purple}{rgb}{1,0,1}

\newcommand{\half}{\frac{1}{2}}
\newcommand{\gsim}{\gtrsim}
\newcommand{\lsim}{\lesssim}
\newcommand{\ra}{\rightarrow}

\newcommand{\ct}{c_\theta}

\def\lag{\mathcal{L}}

\def\shatsq{\sqrt{s}}

\newcommand{\beq}{\begin{equation}}
\newcommand{\eeq}{\end{equation}}
\newcommand{\bea}{\begin{eqnarray}}
\newcommand{\eea}{\end{eqnarray}}
\newcommand{\nn}{\nonumber}

\def\mlq{m_{\rm LQ}}
\def\wlq{\Gamma_{\rm LQ}}
\def\ace{A_{\rm CE}}
\def\afb{A_{\rm FB}}
\def\afbpart{A_{\rm FB}^{\rm CM}}

\def\afbexp{A_{\rm FB}^{\rm CS}}

\def\thcs{\theta_{\rm CS}}
\def\mll{m_{\ell \ell}}
\def\mee{m_{ee}}
\def\mmm{m_{\mu \mu}}
\def\yql{y_{q\ell}}
\def\yue{y_{ue}}
\def\yum{y_{u\mu}}
\def\yde{y_{de}}
\def\ydm{y_{d\mu}}

\allowdisplaybreaks

\begin{document}	

\title{Anticipating Non-Resonant New Physics in Dilepton Angular Spectra at the LHC}

\author{Nirmal Raj}

\affiliation{Department of Physics, University of Notre Dame, 225 Nieuwland Hall, Notre Dame, Indiana 46556, USA}

\begin{abstract}

At the LHC, dileptonic events may turn up new physics interacting with quarks and leptons.
The poster child for this scenario is a resonant $Z'$, much anticipated in $\ell^+ \ell^-$ invariant mass spectra.
However, {\em angular} spectra of dileptons may play an equal or stronger role in discovering non-resonant species.
This paper avails of their LHC measurements to corner the couplings and masses of leptoquarks (LQs), that can mediate $q \bar{q} \ra \ell^+ \ell^-$ in the $t$-channel and dramatically alter Standard Model angular spectra.
Also derived are constraints from alterations to $\mll$ distributions. 
These dilepton probes, exploiting the high rates and small uncertainties of the Drell-Yan process, rival or outdo dedicated LHC searches for LQs in single and pair production modes.
The couplings of LQs with electronic interactions are best bound today by low-energy measurements of atomic parity violation, but can be probed better by $\ell^+ \ell^-$ measurements in the high luminosity runs of the LHC, with the angular spectra leading the way.
This work also advocates the experimental presentation of boost-invariant angular asymmetries that vanish in the SM.
\end{abstract}

\maketitle


\section{Introduction}

The LHC has commenced Run 2, driving the energy frontier onward.
Leading this frontier is dilepton production, a channel that has fetched particle physics
historic triumphs.
This success may be imputed to its cleanness: dileptons are more precisely reconstructed than most other final states (such as jets + $X$), are easily triggerable, and have backgrounds so well understood and high in rates as to minimize theoretical and statistical uncertainties.
The principle behind discoveries in this channel is simple at its core: 
amplitudes with new states mediating the process add themselves to background amplitudes;
the patterns in which dilepton events are distributed in phase space are altered; 
in these new patterns we distinguish signal features.
The Standard Model (SM) provides a classic example of this phenomenon in the neutral current Drell-Yan (DY) process.
The amplitude involving an $s$-channel $Z$ boson supplies its own contribution to the cross-section, as well as interfering with the photon-mediated amplitude.
The net effect spectacularly modifies $\ell^+ \ell^-$ distributions.
In the dilepton invariant mass ($\mll$) spectrum, it produces an unmistakable peak in the form of a Breit-Wigner resonance.
In the angular spectrum, it produces a left-right or forward-backward asymmetry due to its chiral couplings with SM fermions.

Hunting for a new particle in this channel is an important programme at the LHC.
The usual sequence of ideas that dominates our thought concerning it is as follows (see \cite{Petriello:2008zr,Diener:2009ee,Chiang:2011kq}).
First we look for a Breit-Wigner/Jacobian peak in the invariant/transverse mass spectrum of the neutral/charged current DY process. 
Its location then gives a clear picture of the particle's mass; its width, if resolvable, may reveal the decay rate.
Once discovered this way, more information such as spin and chiral couplings can be extracted from the angular spectrum.
These properties may sift out the ultraviolet physics that gave rise to the resonance.

This sequence, however, in attaching more prominence to the kinematic than angular spectrum, can be problematic for two reasons:

1, in general, there is no guarantee that kinematic spectra will precede angular spectra as harbingers of new physics\footnote{See, e.g., \cite{Rutherford:1911zz}.}.
In fact, the sequence in which the $Z$ boson came to our colliders was quite the reverse!
In $e^+ e^-$ collisions at 30 GeV $\lsim \sqrt{s} \lsim$ 40 GeV, the PETRA experiment first found a non-zero dimuon forward-backward asymmetry ($\afb$) due to weak-electromagnetic interference.
From these measurements was derived a bound: $M_Z \leq 100$ GeV.
Only years later did the Super Proton Synchrotron (SPS), in $p$-$\bar{p}$ collisions, achieve the energies required to produce the $Z$ on-shell and announce a resonant peak. 
(The SPS had by then already discovered the $W$ in this fashion.)
Post-discovery, its spin and chirality properties were disentangled with greater precision. 
The past decade too has seen instances of tantalizing hints in angular distributions. 
The (in)famous excess in the $t$-$\bar{t}$ forward-backward asymmetry at the Tevatron \cite{Aaltonen:2011kc,Abazov:2011rq} inspired a model-building flurry, put to rest at the LHC by the measurement of the charge asymmetry \cite{Chatrchyan:2011hk,ATLAS:2012an}.
More recently, physicists at MTA Atomki have observed a bump 
in the spectrum of opening angles in the $e^+e^-$ decay mode of an excited Be-8 state \cite{Krasznahorkay:2015iga}, catching the attention of model-builders. 
 
2, if new physics is {\em non}-resonant, such as when the mediation is not $s$-channel, no clear peak in kinematic spectra is produced from which information on mass can be determined.
In that case, we have no clear guide as to where to anticipate signals, or with what sensitivity.
Very likely, one may need both kinematic and angular spectra to extract all at once the mass, spin and coupling properties.
Could the first signals arrive in the $\ell^+ \ell^-$ angular distributions?

The main purpose of this work is to demonstrate, with an explicit model, that they indeed could. 
In this work, the exotic of choice is the leptoquark (LQ). 
An LQ carries both colour and lepton number, and can be exchanged in the $t$-channel of the DY process $q \bar{q} \ra \ell^+ \ell^-$.
This has a significant effect on the process.
Due to the addition of a new mode, there is of course a non-trivial modification to production rates.
Less obvious is the effect on production angles.
In general, the angular spectrum is picked by the Wigner $d$ functions.
Their application to $s$-channel exchange is straightforward.
When a spin-1 $Z$ (or $Z'$) mediates in the $s$-channel, we have
\begin{equation}
\frac{d\sigma}{d \Omega} \propto (1+\cos^2\theta) + a \cos\theta,
\label{eq:schangspec}
\end{equation}
where $\theta$ is the angle between the outgoing lepton and incoming quark in the centre-of-momentum frame.
As known well, parity violations then leave their imprint in the angular spectrum by making $a \neq 0$, consequently measured as a left-right or forward-backward asymmetry.
In contrast, a spin-0 state exchanged in the $s$-channel leaves the spectrum flat: the events are isotropic.

The case of $t$-channel exchange is subtler.
To begin with, we know that a spin-1 exchange process will always produce anisotropic events. 
We have seen this in the $t$-channel piece of Bhabha scattering: 
\begin{equation}
\frac{d\sigma}{d \Omega} \propto \frac{1+\cos^4(\theta/2)}{\sin^4(\theta/2)}.
\label{eq:tchangspecV}
\end{equation}
In the case of spin-0 $t$-channel exchange, the events are isotropic only if the mediator is massless.
This is at variance with spin-0 $s$-channel exchange, where isotropy is guaranteed regardless of mediator mass. 
With a non-zero mass $\bar{m}_t$, the $t$-channel angular spectrum is
\begin{equation}
\frac{d\sigma}{d \Omega} \propto \frac{\sin^4(\theta/2)}{(s \sin^2(\theta/2) +\bar{m}^2_t)^2},
\label{eq:tchangspec}
\end{equation}
producing anisotropic events.
This paper deals with massive scalar leptoquarks, with a dilepton angular distribution dictated by Eq.~\ref{eq:tchangspec}, which is qualitatively different from the SM distribution in Eq.~\ref{eq:schangspec}. 
Due to interference with the SM process, the full angular spectrum will be some combination of Eqs.~\ref{eq:schangspec} and \ref{eq:tchangspec}.

\begin{figure*}
\begin{center}
\includegraphics[width=15cm]{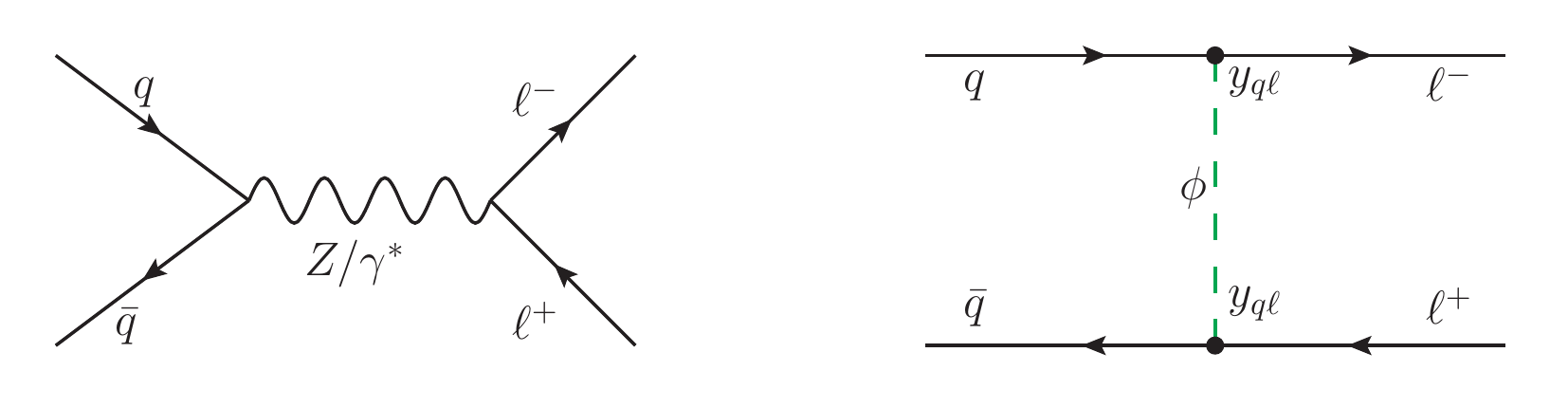} 
\caption{Feynman diagrams that contribute to the Drell-Yan process at tree level.
The $t$-channel leptoquark exchange amplitude on the right
modifies dilepton production by the Standard Model $s$-channel $Z/\gamma^*$ exchange amplitude on the left.
These modifications, trackable in $\ell^+\ell^-$ kinematic and angular spectra, are powerful indirect signals of leptoquarks at the LHC.  
}
\label{fig:feyndy}
\end{center}
\end{figure*}

LQs interacting with first generation quarks are constrained by direct searches at the LHC in processes of pair production \cite{Khachatryan:2015vaa,Aad:2015caa} and single production \cite{Khachatryan:2015qda}.
I will show in this paper that measurements of $\mll$ and the angular dependence (specifically, $\afb$) of $p p \ra \ell^+ \ell^-$, by virtue of their cleanness, provide competitive or stronger limits than these dedicated searches.
In the case of an LQ coupling to electrons, precision low-energy experiments measuring atomic parity violation provide more stringent constraints than $p p \ra  \ell^+ \ell^-$ measurements. 
However, I will show that with the projected high luminosities of the future LHC, the dilepton probes could achieve enough precision to overtake these experiments, with the angular spectrum measurement marking the trail. 

Aspects of this work have appeared in the literature.
The use of $\afb$ was briefly explored in \cite{Hewett:1997ce} to probe a 200 GeV-heavy scalar LQ 
at the Tevatron.
Signals of vector LQs in the $\tau^+\tau^-$ charge asymmetry are shown for the LHC at 14 TeV in \cite{1301.4214}.
In \cite{Accomando:2015cfa},
a sensitivity study for the 13 TeV LHC is performed regarding the use of $\afb$ as the discovery mode of a $Z'$.
More discussion on this study is relegated to Sec.~\ref{sec:discs}. 
Ref.~\cite{1511.04573} estimates the couplings of a $Z'$ boson using LHC Run 1 measurements of the $\afb$.
Ref.~\cite{Altmannshofer:2014cla} investigates
threshold effects from loop processes involving a dark sector that give rise to unique features in dilepton spectra, which can constrain dark matter masses and couplings.
The sizable impact on $\ell^+ \ell^-$ angular spectra was mentioned, but no limits were set.
Angular spectra of jets and top quarks, limited as they are by larger uncertainties than $\ell^+\ell^-$, may nonetheless usher in new physics:
Ref.~\cite{Cerrito:2016qig} studies a 3 TeV-heavy $Z'$ boson discoverable in $t$-$\bar{t}$ angular asymmetries.

In \cite{Wise:2014oea}, LQs were bounded with $e^+ e^-$ and $\mu ^+ \mu^-$ mass spectra at the LHC.
This analysis was performed using Poisson statistics in high $\mll$ bins where no events were observed and the expected SM background was low\footnote{I would like to thank Yue Zhang for sharing this information in private correspondence.}. 
Angular distributions were not considered.
One avenue to probe hidden sectors 
that are not necessarily resonant is to characterize their mediation as contact operators;
ATLAS has set bounds on their size using both $\mll$ spectra and $\afb$ \cite{Aad:2014wca}.
Ref.~\cite{Bessaa:2014jya} recast these bounds at $\sqrt{s} = 7$ TeV to constrain LQs.
The LQ species here differs from the ones used in \cite{Wise:2014oea} (and this work).
Nevertheless, the authors claim that their bounds are about 25\% weaker than those set by \cite{Wise:2014oea}.
Ref.~\cite{Bessaa:2014jya} also makes the important clarification that it is not possible to recast the ATLAS bounds for LQ species giving rise to multiple operators.
In \cite{deBlas:2013qqa}, a general parametrization capturing the combination of {\em all} quark-lepton contact operators was presented.
It was shown how, with the help of this parametrization, an $\afb$ measurement at $\sqrt{s} = 14$ TeV could identify various operator combinations.
Operator analyses such as \cite{Aad:2014wca,Bessaa:2014jya,deBlas:2013qqa} neglect, by construction, the momentum dependence in the LQ propagator; in contrast, this work accounts for the full propagator of LQ-mediated processes. 

The $\mll$ spectrum may be sensitive to renormalization group (RG) running of electroweak (EW) couplings \cite{Rainwater:2007qa,Alves:2014cda,Gross:2016ioi} and EW precision test parameters \cite{1609.08157}. 
Effects of $t$-channel mediation
are explored in \cite{Nie:2000kz,Cheng:2014rba,Jelinski:2015epa}.

This paper is set up as follows.
Sec.~\ref{sec:LQs} reviews LQs, selects models for study and introduces some useful terminology.
Sec.~\ref{sec:probes} discusses in detail how LHC measurements of dileptonic $\mll$ spectra and $\afb$ can be used to set limits on LQs.
Using LQs for illustration, it also promotes the depiction of angular spectra with the centre-edge asymmetry, a boost-invariant observable that could vanish in the SM.
Sec.~\ref{sec:otherprobes} briefly reviews conventional probes of LQs, consigning to an appendix the elaboration of methods used to recast observed bounds.
Sec.~\ref{sec:results} presents the results of the previous two sections and forecasts the sensitivity of dilepton probes to LQs at future LHC runs.
Sec.~\ref{sec:discs} summarizes the paper, and discusses its scope and related future work.

\section{Leptoquark Models}
\label{sec:LQs}

Leptoquarks (LQs) are exotic particles having both baryon and lepton number, carrying such quantum numbers and spins as to mediate interactions between quarks and leptons through a renormalizable vertex, schematically given by 
\begin{equation*}
\lag \supset {\rm (lepton)(LQ)(quark)}.
\end{equation*}
For a comprehensive review, see \cite{Dorsner:2016wpm}. 

LQs are an ill-studied breed in the trade, for their inability to pose as ready solutions to current problems. 
Nonetheless, the case made for their existence is this: 

(i) They appear as infrared remnants of grand unified theories \cite{Gershtein:1999gp,Dorsner:2005fq}.

(ii) They may be the mediators of dark matter-SM interactions \cite{Baker:2015qna}.

(iii) They feature in some technicolour and composite models \cite{Dimopoulos:1979es,Farhi:1980xs,Schrempp:1984nj}.

(iv) They appear in $R$-parity violating versions of supersymmetry \cite{Barbier:2004ez}.

(v) They may explain a number of anomalies in low-energy flavour experiments \cite{Dorsner:2016wpm}.

(vi) If one looks in colliders for new physics at the TeV scale that is in discoverable form, the corresponding beyond-the-SM particles are likely to have renormalizable interactions with at least one SM fermion. 
Among scalars, the possibilities are: colour singlets or octets with Higgs-like EW charges, mediators of singlet fermion-SM fermion interactions (such as sfermions), ``diquarks", ``dileptons" and lastly, leptoquarks \cite{Giudice:2011ak}.

This work is entirely in line with the last of these motivations.

Depending on their gauge charges and Lorentz structure, several species of LQs are possible.
These models are enumerated in \cite{Dorsner:2016wpm}.  
In this work I will confine myself to scalar LQs. 
A brief discussion of vector LQs 
is given in Sec.~\ref{sec:discs}.

There are four scalar LQ species that violate baryon number and two that do not.
To avoid dealing with constraints from rapid proton decays, I will only treat the latter two.
In the notation of \cite{Dorsner:2016wpm}, these are $R_2 ({\bf 3,2},7/6)$ and $\tilde{R_2}({\bf 3,2},1/6)$, where the quantities in parantheses denote the $SU(3)_c \otimes SU(2)_W \otimes U(1)_Y$ quantum numbers.
The Lagrangian involving the first of these is given by\footnote{Also present are quartic terms involving the Higgs field, $\lambda|H|^2|{\rm LQ}|^2$, and QCD interactions. 
These may modify SM production rates and branching ratios of the Higgs boson through loops and confront limits from the corresponding LHC measurements \cite{Dorsner:2016wpm}.
These limits do not affect the phenomenology of this work.
The Higgs-LQ quartic term introduces another constraint.
As the LQ species here are $SU(2)_W$ doublets, this term induces a mass splitting between their components through EW symmetry breaking, $\Delta m = - \lambda v^2/\mlq$.
This results in new contributions to the oblique parameters $S$ and $T$, constrained by precision EW data.
The 95\% C.L. bound on the splitting for $\Delta m \ll \mlq$ is $\Delta m \leq 53$ GeV \cite{Dorsner:2016wpm}, which is viable at small $\lambda$ and/or high $\mlq$.}

\beq
\nn \lag =  - y_{ij} \bar{u}_{R,i} R_2^a \epsilon^{ab} L_{L,j}^b + y'_{ij} \bar{e}_{R,i} {R^a_2}^* Q^a_{L,j} + {\rm h.c.}
\eeq

In the mass basis, this becomes
\bea
\nn \lag &=&  (y V_{\rm PMNS})_{ij}  \bar{u}_{R,i} \nu_{L,j} R_2^{2/3} - y_{ij}  \bar{u}_{R,i} e_{L,j} R_2^{5/3} \\ 
\nn &+& y'_{ij}  \bar{e}_{R,i} d_{L,j} {R_2^{2/3}}^*  + (y' V^\dagger_{\rm CKM})_{ij}  \bar{e}_{R,i} u_{L,j}{R_2^{5/3}}^* + {\rm h.c.} \\
 \label{eq:Lag1}
 \eea

The indices $i$ and $j$ run over fermion families.
In order to make direct comparisons to results in the literature, I now choose a flavour structure that is in vogue.
Following \cite{Wise:2014oea}, I set $y'_{ij} = 0$ and consider two possibilities for the matrix $y_{ij}$: the up quark is invited to couple either to the electron family or the muon family, i.e., either $y_{ij} = \yue \delta_{i1}\delta_{j1}$, or $y_{ij} = \yum \delta_{i1}\delta_{j2}$.
One virtue of this choice is that a vanishing $y'_{ij}$ allows one to separate the LQ's couplings to up-type quarks from down-type, simplifying their treatment.
Another important virtue will be seen in Sec.~\ref{sec:results}.
When an LQ mediates valence quark-electron interactions, the low-energy measurement of atomic parity violation (APV) easily outconstrains LHC searches, rendering them redundant.
However, the LHC lamppost provides the best limits on LQs mediating valence quark-muon interactions since no APV measurement on muonic systems has been performed as of now.
LQ couplings to other second and third quark generations are not considered here since the DY process at a $p$-$p$ collider, the focus of this paper, proceeds effectively through valence quarks.
Couplings to the tau lepton are also not considered.

\begin{figure*}
\begin{center}
\includegraphics[width=.45\textwidth]{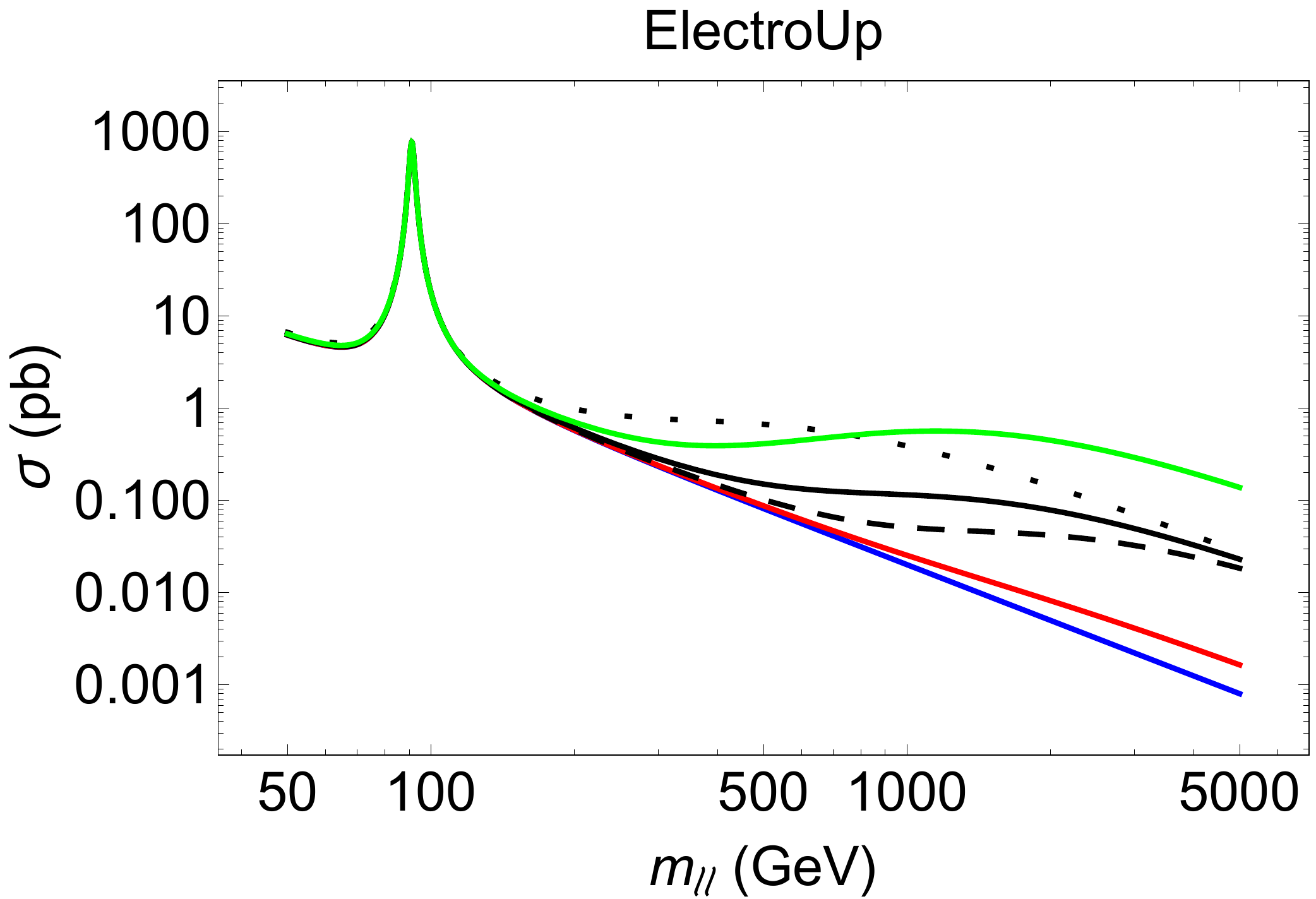} 
\quad \quad
\includegraphics[width=.45\textwidth]{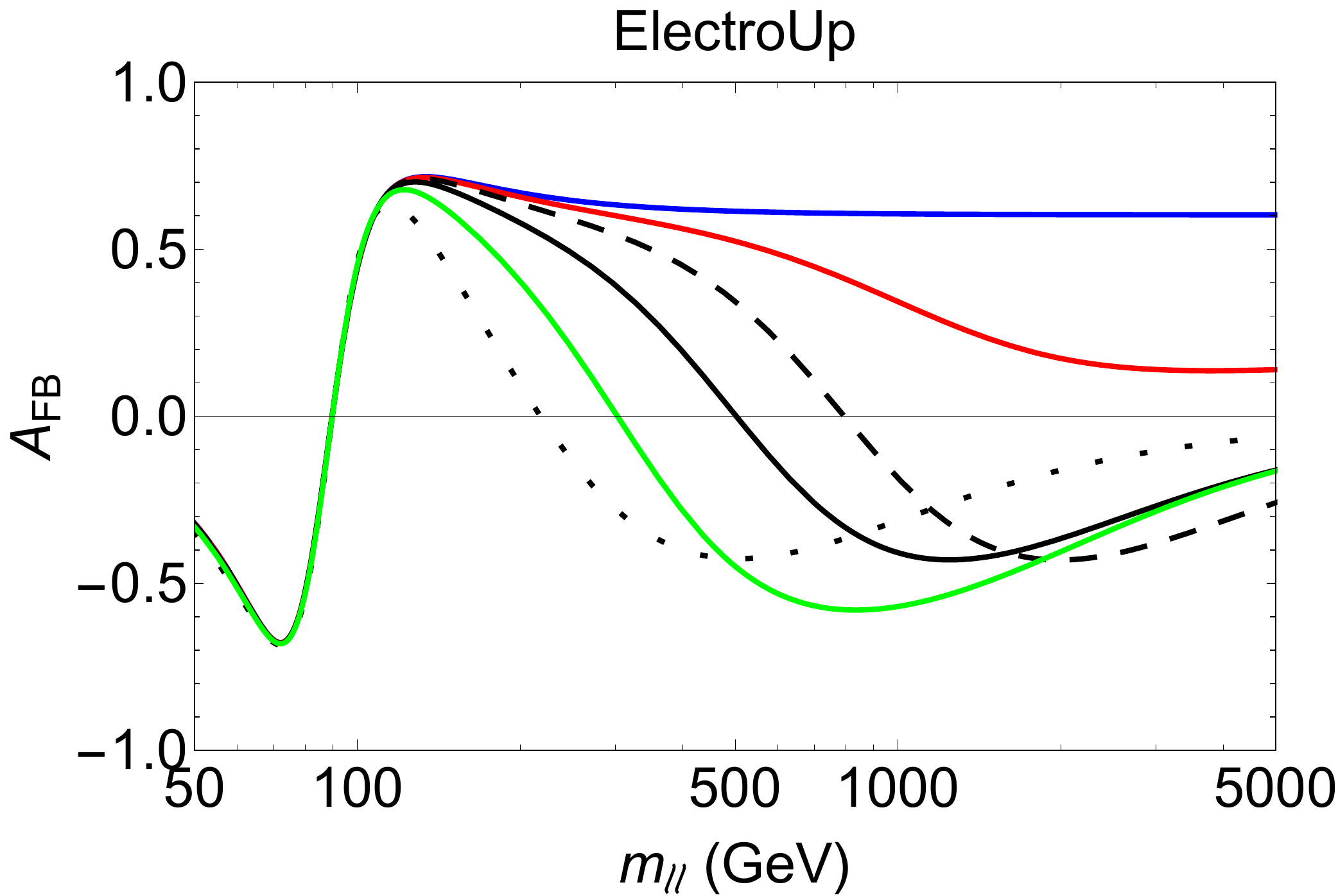} 
\\
\vspace{1cm}
\includegraphics[width=.45\textwidth]{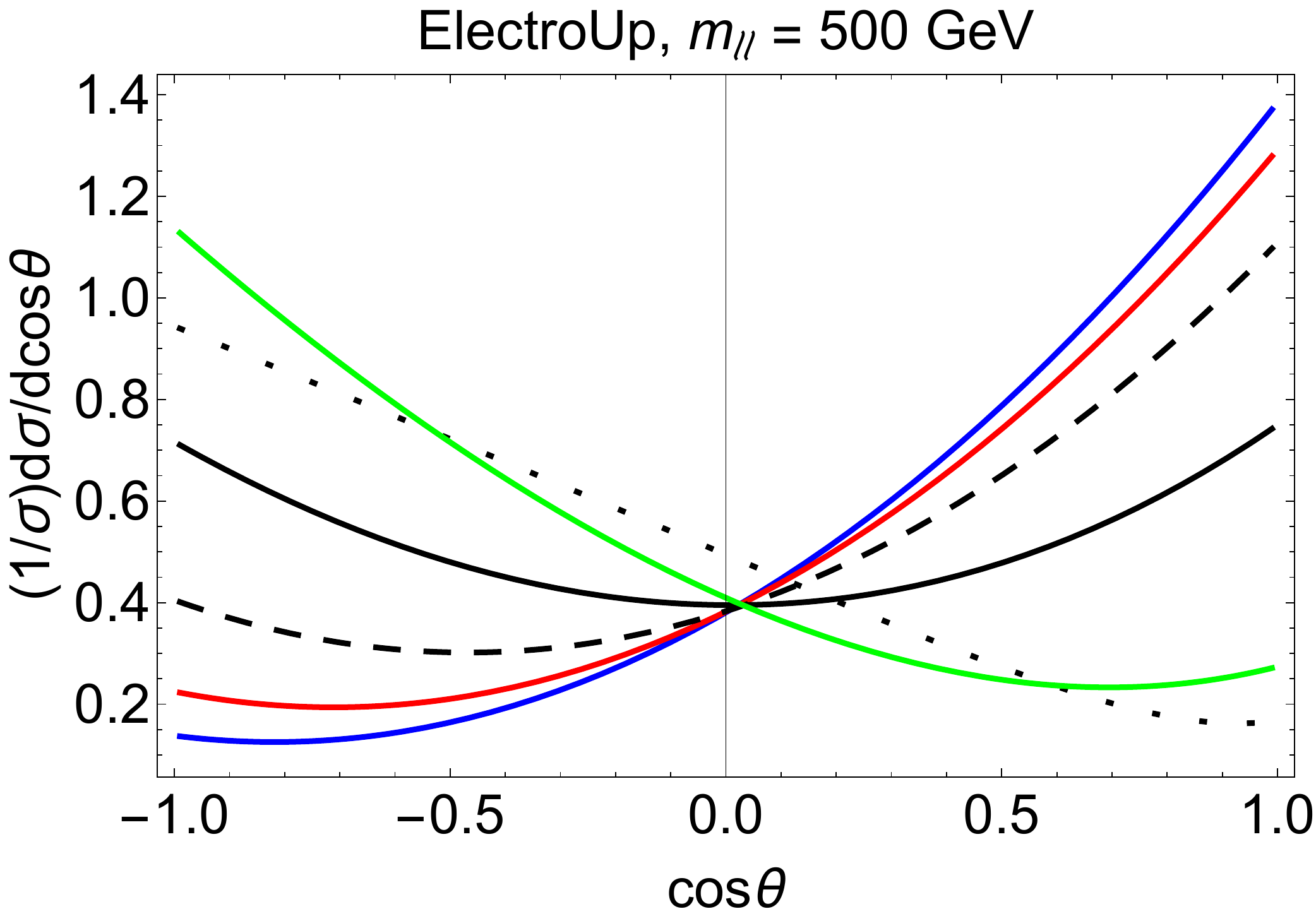}
\quad \quad
\includegraphics[width=.45\textwidth]{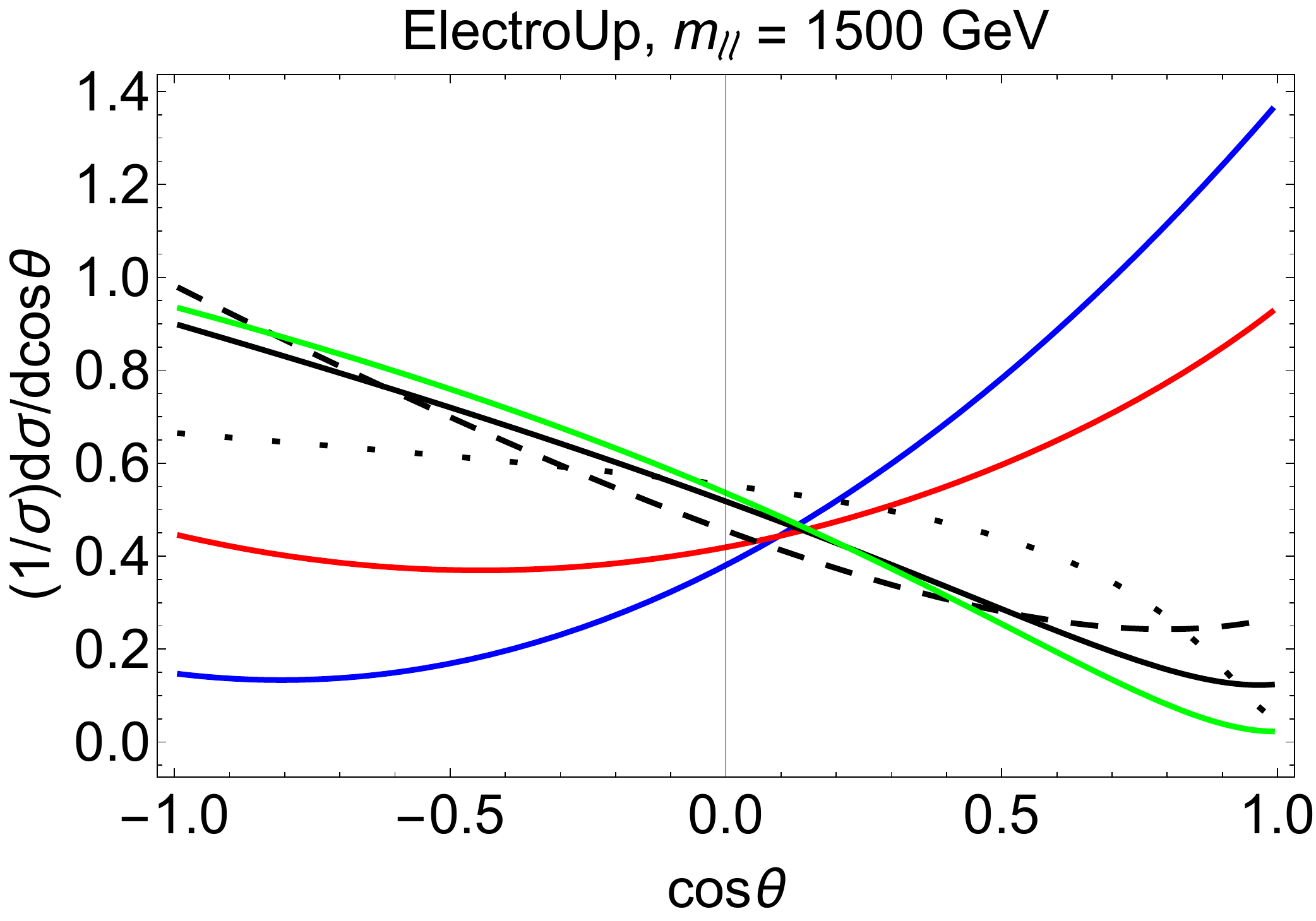}
\caption{Effect of leptoquarks on dielectron production spectra, depicted at the partonic level.
An {\tt ElectroUp} is chosen for illustration.
As a function of $\mll$, the top left plot shows the production cross-section  and the top right plot the parton level forward-backward asymmetry defined in Eq.~\ref{eq:afbdef}.
The bottom plots show the normalized angular distributions at $\mll =$ 500 GeV and 1500 GeV.
Here the blue curves are obtained from the $Z/\gamma^*$-mediated diagrams in Fig.~\ref{fig:feyndy} with an up quark-antiquark initial state.
The solid curves show deviations from the SM spectrum as $\mlq$ is kept fixed at 1 TeV and the LQ Yukawa $\yue$ is varied, with \{red, black, green\}: \{0.4, 1, 1.6\}.
The black curves show deviations as $\yue$ is fixed at 1 and $\mlq$ is varied, with \{dotted, solid, dashed\}: \{400, 1000, 1600\} GeV.
More details are described in the text.
}
\label{fig:deviations}
\end{center}
\end{figure*}

The interactions of the second LQ species are given by
\bea
\nn \lag &=& - y_{ij}  \bar{d}_{R,i} \tilde{R}_2^a \epsilon^{ab} L_{L,j}^b + {\rm h.c.} \\
\nn &=& - y_{ij}  \bar{d}_{R,i} e_{L,j} \tilde{R}_2^{2/3} + (y V_{\rm PMNS})_{ij}  \bar{d}_{R,i} \nu_{L,j}\tilde{R}_2^{-1/3} + {\rm h.c.} \\
\label{eq:Lag2}
\eea

In the second line, I have rewritten the first line in the mass basis.
I choose now a flavour structure similar to the one just discussed: the down quark is made to communicate exclusively to either the electron family or the muon family.
Thus, either $y_{ij} = \yde \delta_{i1}\delta_{j1}$, or $y_{ij} = \ydm \delta_{i1}\delta_{j2}$.

Some notes are in order {\em re} the Yukawa coupling structures imposed in Eqs.~\ref{eq:Lag1} and \ref{eq:Lag2}.
First, one presumes that these structures are imposed at some high scale by strange symmetries in the diagonal basis of SM Yukawa couplings.
In that case, RG running 
may introduce non-zero off-diagonal elements in all these matrices.
When diagonalized to the fermion mass basis, one may wonder if unacceptably large flavour-changing neutral currents (FCNCs) are induced.
However, there is little cause for such worry: it was shown in \cite{Wise:2014oea} that, even if these coupling structures are set at a scale as high as $M_{\rm {Planck}}$, the models are safe from FCNC constraints.
This is due to suppression from down-type SM Yukawa strengths and loop factors that appear in the RG running.

Second, in both Eqs.~\ref{eq:Lag1} and \ref{eq:Lag2}, the choice of our coupling matrices gives LQs mediating quark-neutrino interactions.
These interactions will be irrelevant to the phenomenological focus of this work, since 
$\afb$ measurements were performed at the LHC only in charged lepton final states.
See also \cite{Wise:2014oea}, where it was shown that constraints from neutrino experiments are weaker than those from charged dilepton production in coupling-mass space.
This work will not be concerned with LQ-neutrino interactions.

In all, I have described four types of leptoquarks above, mediating interactions between four distinct quark-lepton combinations.
I would like to distinguish among these with a terminology that clarifies exactly which lepton-quark combination is at play.
Thus I will refer to the two $R_2^{5/3}$ LQs as 
{\tt ElectroUp}
and
{\tt MuoUp},
and to the two $\tilde{R}_2^{2/3}$ LQs
as {\tt ElectroDown}
and
{\tt MuoDown}.
On occasion, I will also use terms such as 
{\tt MuoQuark},
{\tt LeptoDown}, etc.
to collectively denote the LQs that interact with the named lepton or quark.
Additionally, I will use the symbol ``$\yql$" when I discuss the LQ Yukawa coupling in a generic manner.

\section{Dilepton Probes} 
\label{sec:probes}
\begin{figure*}[!htbp]
\begin{center}
\includegraphics[width=.45\textwidth]{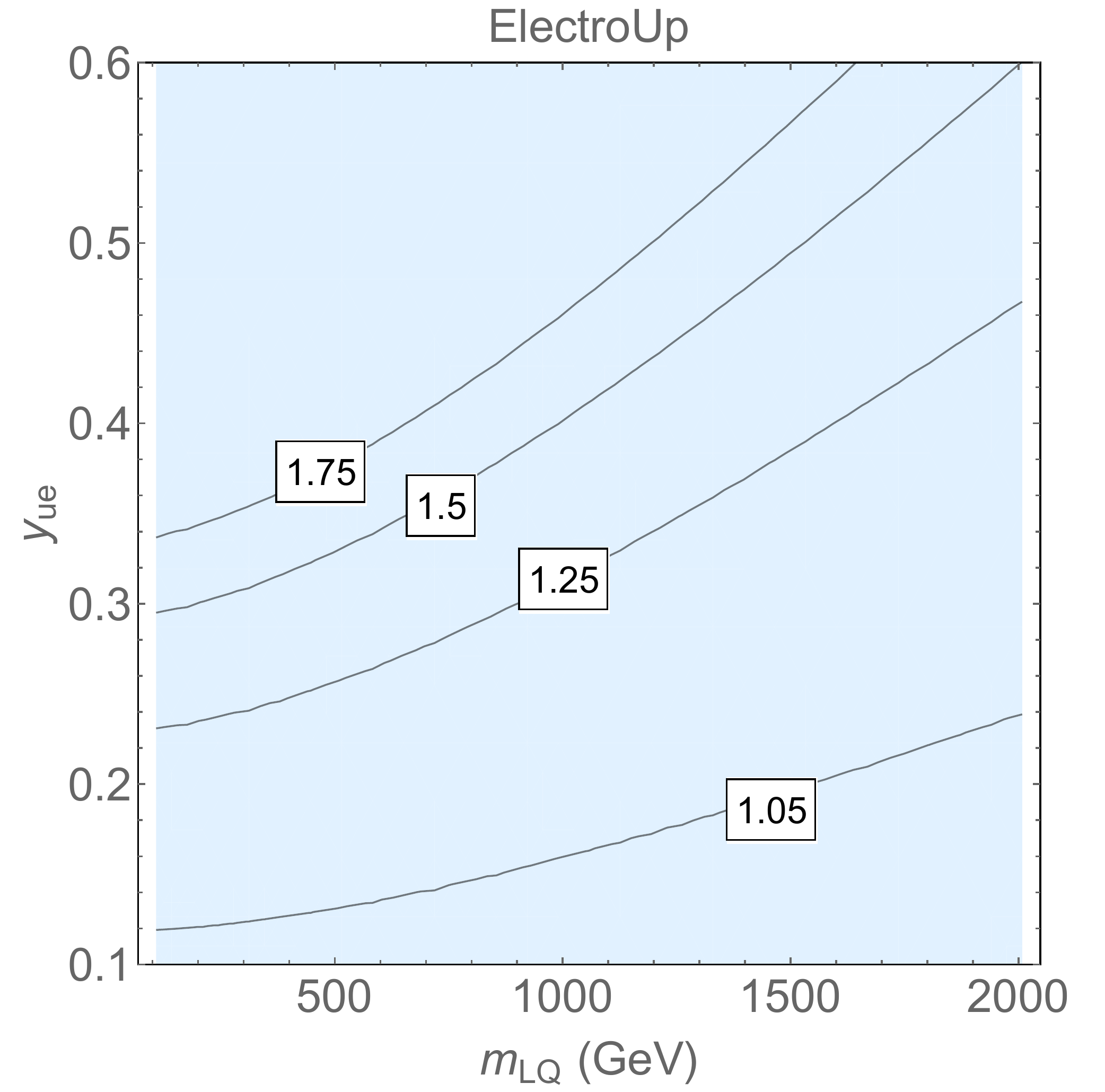} 
\quad \quad
\vspace{1cm}
\includegraphics[width=.45\textwidth]{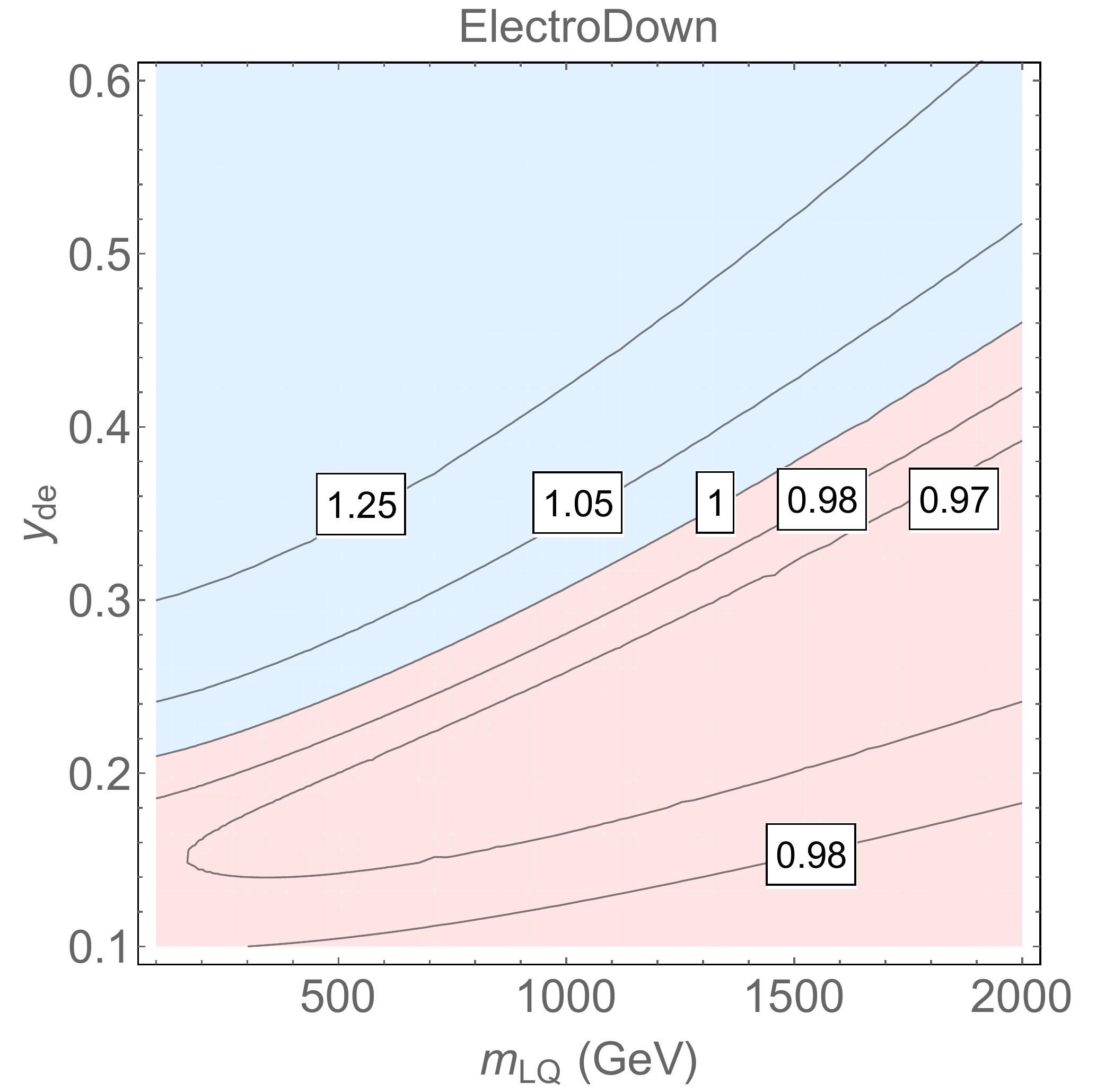} 
\caption{Contours of the ratio $d\sigma_{\rm tot}/d\sigma_{\rm SM}$ in the {\tt ElectroUp} (left) and {\tt ElectroDown} models (right) at $\mll = 1500$ GeV.
Regions where the ratio $> 1 \ (< 1)$ are shaded blue (red).
This illustrates that destructive interference between SM and leptoquark-mediated Drell-Yan production (i.e., between the diagrams in Fig.~\ref{fig:feyndy}) is possible for {\tt LeptoDowns}.
Since the up quark is denser in protons than the down quark, the dip in hadronic level cross-sections is more subdued than at the partonic level.
See text for more details.}
\label{fig:sinks}
\end{center}
\end{figure*}

This section describes the effect of LQs on dilepton events at the LHC.
It begins with the $\mll$ distribution, which may be more familiar to the reader.
Next addressed, in three sub-sections, are angular spectra.
First described is the forward-backward asymmetry in the Collins-Soper reference frame.
Next discussed are signals visible in asymmetries built from frame-independent quantities.
Finally discussed, briefly, is the ATLAS measurement of angular distributions in the Collins-Soper frame. 

This section also details how LHC dilepton measurements can be recast to restrict LQ parameters. 

\subsection{$\mll$ distributions}
\label{sec:mll}

Let me begin with an overview of the effect that $t$-channel LQ mediation  has on the $\mll$ spectrum.
Denoting by $\theta$ the angle between the incoming quark and the outgoing lepton in the centre-of-momentum frame, and taking quarks and leptons massless, the parton level differential cross-section for the process
$q \bar{q} \ra \ell^+ \ell^-$ at leading order (LO) is given by
\bea
\nn d\sigma_{\rm tot} &\equiv&  \frac{d\sigma_{\rm tot}}{d\ct}  \\
 &=& d\sigma_{\rm SM} + d\sigma_{\rm int} + d\sigma_{\rm LQ}~,
\label{eq:XSexpansion}
\eea
with
\bea
\nn d\sigma_{\rm SM} &=& \frac{1}{32 \pi \mll^2 N_c} \sum_{\rm spins} |\mathcal{M_{\rm SM}}|^2~, \\
\nn d\sigma_{\rm int} &=& -\frac{1}{32 \pi \mll^2 N_c} \sum_{\rm spins} 2{\rm Re}(\mathcal{M_{\rm SM}}\mathcal{M_{\rm LQ}^*})~, \\
d\sigma_{\rm LQ} &=& \frac{1}{32 \pi \mll^2 N_c} \sum_{\rm spins} |\mathcal{M_{\rm LQ}}|^2~,
\label{eq:XSdefs}
\eea

where $N_C = 3$ is the number of QCD colours, $\mathcal{M_{\rm SM}} = \mathcal{M_{\gamma}} + \mathcal{M_{ Z}}$ is the SM amplitude corresponding to the Feynman diagram on the left of Fig.~\ref{fig:feyndy} and $\mathcal{M_{\rm LQ}}$ is the amplitude for the LQ-mediated diagram on the right of Fig.~\ref{fig:feyndy}.
The latter are given by 
\bea
\nn \mathcal{M_{ \gamma}} &=& i Q_q e^2 [\bar{v}(p_{\bar{q}})\gamma^\mu u(p_{{q}})] \frac{-g_{\mu\nu}}{\mll^2} [\bar{u}(p_{\ell^-})\gamma^\nu v(p_{\ell^+})]~, \\
\nn \mathcal{M_{ Z}} &=& i [\bar{v}(p_{\bar{q}})\gamma^\mu (g_L^q P_L + g_R^q P_R)u(p_{{q}})] \frac{-g_{\mu\nu}}{\mll^2-M_Z^2 - i \Gamma_Z M_Z} \\
\nn & & \ [\bar{u}(p_{\ell^-})\gamma^\nu(g_L^q P_L + g_R^q P_R) v(p_{\ell^+})]~, \\
\mathcal{M_{\rm LQ}} &=& i \yql^2 [\bar{v}(p_{\bar{q}}) P_R v(p_{{\ell^+}})] \frac{1}{\hat{t}-\mlq^2}[\bar{u}(p_{\ell^-}) P_L u(p_{{q}})] ~.
\label{eq:ampdefs}
\eea
Here $e$ is the electromagnetic coupling, $Q_q$ is the quark electric charge, $\Gamma_Z$ and $M_Z$ respectively the decay width and mass of the $Z$ boson, with couplings to the fermions $f$ given by
 $g^f = (e/\cos\theta_W \sin \theta_W) (T^f_3 - Q^f \sin^2\theta_W)$.
 The minus sign in the interference term in Eq.~\ref{eq:XSdefs} comes from the relative ordering of external spinors seen in Eq.~\ref{eq:ampdefs}.

It is not immediately obvious whether $\mathcal{M_{\rm LQ}}$ interferes with $\mathcal{M_{\rm SM}}$ constructively or destructively, a question to which I will return shortly.
What {\em is} clear is that if constructive interference transpires, $\ell^+ \ell^-$ production rates increase with $\yql$.
Moreover, at $\mll \gg \mlq$, one expects the total dilepton cross-section $d\sigma_{\rm tot}$ to be offset from  the SM cross-section $d\sigma_{\rm SM}$ by a more-or-less constant factor:
in this region the LQ plays a massless mediator contributing an extra channel to the rate of $\ell^+\ell^-$ production\footnote{Contrast this with \cite{Bessaa:2014jya}, where the region under consideration is $\mll \ll \mlq$ so that the amplitude $\mathcal{M_{\rm LQ}}$ can be written as a contact interaction.}.

These traits are seen in the plot on the left-hand side of Fig.~\ref{fig:deviations}, where I have illustrated them using an {\tt ElectroUp} with various masses and couplings.
These curves denote $d\sigma_{\rm tot}$ integrated over $\cos\theta$.
The blue curve corresponds to $\yue = 0$, viz., $d\sigma_{\rm tot} \ra d\sigma_{\rm SM}$.
The solid curves demonstrate the effect of LQ mediation when $\mlq$ is fixed and $\yue$ is varied.
Here I keep $\mlq = 1$ TeV, and denote by the red, black and green curves  $\yue = 0.4, 1$ and 1.6 respectively.
The rates patently rise with the coupling.
The black curves show how the effect changes with $\mlq$ keeping $\yql$ fixed.
At $\yue = 1$, the dotted, solid and dashed curves represent $\mlq = 400, 1000$ and 1600 GeV respectively.
After rising across $\mll$, the dotted curve flattens to a value that is at a constant offset from the blue curve.
All three black curves are seen to asymptote to this value.

Are there destructive interferences?
Among the LQ models considered here, it is possible to have them when the mediation is provided by a {\tt LeptoDown}, while {\tt LeptoUps} always interfere constructively with the SM DY process. 
This is because of electric charge, as seen in Eq.~\ref{eq:ampdefs}.
In Fig.~\ref{fig:sinks}, this effect is illustrated with contours of the ratio $r \equiv d\sigma_{\rm total}/d\sigma_{\rm SM}$ in the $\yue$-$\mlq$ plane, using the {\tt ElectroUp} and {\tt ElectroDown} models as examples. 
The blue (red) regions correspond to $ r > 1  \ (r < 1)$.
The {\tt ElectroUp} always gives $r >1$. 
The {\tt LeptoUp} gives $r<1$ when $\yde$ is small and/or $\mlq$ is high, which is the region where LQ-SM interference ($d\sigma_{\rm int}$) dominates the cross-section.
As $\yde$ is increased and $\mlq$ lowered, the LQ-LQ contribution ($d\sigma_{\rm LQ}$) dominates to make $r > 1$.
The effect of the {\tt LeptoUp}'s destructive interference will be less dramatic at hadronic level cross-sections, due to the smaller down quark densities in the proton than the up quark.

At the LHC, dilepton production rates were measured at 8 TeV with an integrated luminosity of about 20 ${\rm fb}^{-1}$.
Events were recorded up to $\mee \sim 1600$~GeV and $\mmm \sim 1800$~GeV by ATLAS \cite{Aad:2014cka} and $\mee \sim 1750$~GeV and $\mmm \sim 1850$~GeV by CMS \cite{Khachatryan:2014fba}.
The dominant and irreducible background is the neutral current DY process, $q \bar{q} \ra \ell^+ \ell^-$, which in the SM proceeds through the diagram on the left-hand side of Fig.~\ref{fig:feyndy}.
The subdominant backgrounds come from production of dibosons, top quarks, dijets and $W$+jets.

The above measurements can constrain LQ parameters.
I remind the reader that these constraints constitute only the secondary result of this paper: the primary result will be the limits from $\ell^+\ell^-$ angular spectrum measurements, with which I deal in the next section.
I will now employ only the ATLAS $\mll$ measurements toward my constraints, for two reasons --
(i)
both collaborations have similar sensitivities and their results concur: the measured data was consistent with the SM. 
Thus both measurements place similar exclusion limits on the leptoquark models;
(ii)
the ATLAS measurement is presented as an event distribution across $\mll$ with the bins evenly spaced on a logarithmic axis. 
CMS presents a distribution of Events/GeV across $\mll$ and the bins are not evenly spaced.
Consequently, the process of determining the exact number of events in each bin is error-prone.

To set constraints using $\mll$ distributions, I repeat the procedure used in \cite{Altmannshofer:2014cla}.
Cross-sections for the process $p p \ra \ell^+ \ell^-$ are obtained analytically by convolving the partonic level processes $q \bar{q} \ra \ell^+ \ell^-$ with MSTW2008NNLO parton distribution functions (PDFs).
The common renormalization and factorization scale is taken as $\mll$.
The SM background is taken to comprise solely of the $s$-channel $Z/\gamma^*$-mediated process on the LHS of Fig.~\ref{fig:feyndy} -- the sub-dominant backgrounds are neglected.
To generate signal events, I take the $\mll$ distribution of the NNLO background provided by ATLAS and scale it by $\mll$ distributions of the ratio $d\sigma_{\rm tot}/d\sigma_{\rm SM}$, where $d\sigma_{\rm tot}$ and $d\sigma_{\rm SM}$ are as defined in Eq.~\ref{eq:XSexpansion}.
It is not unreasonable that most of the next-to-leading (NLO) order corrections from QCD and electro-weak effects are common to both the signal and background, and would thus disappear in the ratio.
PDF uncertainties similarly fall out.
No RG-improvement on the couplings is performed as it has no significant impact on the final results.
Finally, the analytical cross-sections are validated with {\tt MadGraph5} \cite{Alwall:2014hca}. 
To do this I used the UFO files generated by \cite{Baker:2015qna}, with suitable modifications to include {\tt MuoQuarks} and {\tt LeptoDowns}. 

In this analysis, I take events far from the $Z$ peak (above $\mll = 500$ GeV) as this is where the leptoquarks of masses considered here have their highest impact.
In these high $\mll$ bins the statistical errors dominate. 
Limits can be obtained from a shape analysis comparing the SM and new physics dilepton distributions. 
One does this by finding $\Delta \chi^2 = \chi^2_{\rm NP} - \chi^2_{\rm SM}$, where
\bea
\nn \chi^2_{\rm NP} &=& \sum_{i = 1}^{N_{\rm bins}} \frac{(N^i_{\rm obs} - N^i_{\rm NP})^2}{N^i_{\rm NP}+\sigma^2_{\rm SM}}, \\
\chi^2_{\rm SM} &=& \sum_{i = 1}^{N_{\rm bins}} \frac{(N^i_{\rm obs} - N^i_{\rm SM})^2}{N^i_{\rm SM}+\sigma^2_{\rm SM}}
\label{eq:chisqmll}
\eea

with $\sigma_{\rm SM}$ the systematic uncertainty of the background.

The results of the above procedure are plotted in Fig.~\ref{fig:limits} with red curves.
Also plotted using magenta curves are the results of Ref.~\cite{Wise:2014oea}, which used a somewhat similar analysis.
Here, only high bins with $\mll \geq 1.8$ TeV were considered, where no events were observed by ATLAS. 
The interference term $d\sigma_{\rm int}$ was neglected, since large couplings were probed.
Limits were then set using Poisson statistics.
All these results are discussed in Sec.~\ref{sec:discs}.

\subsection{Dilepton angular distributions}
\label{sec:angspec}

Defining $\ct \equiv \cos\theta$, where $\theta$ is given in Eq.~\ref{eq:schangspec},
the angular distribution of dilepton production rates can be written as
\begin{equation}
 \frac{d^2\sigma}{d\mll d\ct} = \sum_{n=0}^{\infty} a_n \ct^n~,~~~~ a_n\in\mathcal{R} ~,
 \label{eq:gendXSdct}
\end{equation}
where the $a_n$ coefficients are $\mll$-dependent.
The bottom plots of Fig.~\ref{fig:deviations} show the dramatic deviations from the SM angular spectrum
caused by leptoquarks.
The colour code here is the same as in the $\mll$ spectra, and now the left-hand (right-hand) plot corresponds to $\mll = 500$ GeV ($\mll$ = 1500 GeV).
As before, an {\tt ElectroUp} is chosen for illustration.
The spectra are normalized with respect to the $\ct$-integrated cross-section.
Due to the qualitative differences seen between the SM and LQ-contributed spectra, one expects LQs to produce considerable departures from the SM in observables that characterize the angular dependencies (such as the forward-backward asymmetry). 
 
Now at a $p$-$p$ collider such as the LHC, it is impossible to determine in each event the origin of the initial state quark or anti-quark, on account of which there exists an inherent uncertainty in the determination of the angle $\theta$.
It is further complicated by uncertainties in the transverse momenta of partons.
These difficulties are partly overcome by following the prescription of Collins and Soper 
(CS) \cite{Collins:1977iv}, in which the lepton scattering angle in question, $\theta_{\rm CS}$, is distinct from $\theta$.
I will expand on the details of this frame in Sec.~\ref{sec:afb}.

A common experimental practice to measure the angular distribution in Eq.~\ref{eq:gendXSdct} is to determine the forward-backward asymmetry. 
In Sec.~\ref{sec:afb}, I will first discuss how this measurement can be used as a probe of leptoquark parameters, and explain the procedure I use to place bounds.
Next, in Sec.~\ref{sec:otherangulars}, I will
show how the centre-edge asymmetry, a boost-invariant observable that can vanish in the SM, can improve on the $\afb$ as a probe of New Physics, and will illustrate the case with LQs.
Finally, in Sec.~\ref{sec:costheta}, I will briefly discuss the dilepton event distributions in $\cos\theta_{CS}$ as measured by ATLAS at $\shatsq = 8$~TeV.
I place no bounds based on this measurement -- as I will explain in that sub-section, modelling the full background is beyond the scope of this paper.  

\subsubsection{Forward-backward asymmetry}
\label{sec:afb}

The forward-backward asymmetry is a conventional variable used to measure the dilepton angular distribution in Eq.~\ref{eq:gendXSdct}.
It is given by
\begin{eqnarray}
 \afb(\mll) &\equiv& \frac{\left[\int_{0}^{1} - \int_{-1}^{0}\right] d\ct(d^2\sigma/d\ct d\mll)}{\left[\int_{0}^{1} + \int_{-1}^{0}\right] d\ct
 (d^2\sigma/ d\ct d\mll)}  \nonumber \\
 &=& \frac{(d\sigma/d\mll)_{\rm F} - (d\sigma/d\mll)_{\rm B}}{(d\sigma/d\mll)_{\text{tot}}} \nn \\
 &=& \frac{N_{\rm F} - N_{\rm B}}{N_{\rm tot}} ~.
\label{eq:afbdef}
\end{eqnarray}

The top right-hand plot of Fig.~\ref{fig:deviations} illustrates the dramatic deviations from the SM $\afb$ at partonic level due to the inclusion of {\tt ElectroUp}-mediated dilepton production.
The colour code follows the other Fig.~\ref{fig:deviations} plots.
The SM curve varies markedly below and near the $Z$ pole due to $Z$-$\gamma^*$ interference.
It changes sign close to the $Z$ pole and at high $\mll$ settles to a steady value near $\sim 0.6$.
We may compute this value analytically as follows.
Using Eqs.~\ref{eq:XSdefs} and \ref{eq:ampdefs} in Eq.~\ref{eq:afbdef}, one finds for $\mll \gg M_Z$

\beq
 \afb (\rm SM) = \frac{3}{4} \cdot \frac{\beta_e \Gamma^1_- + \Gamma^2_-}{\beta_e^2 + \beta_e \Gamma^1_+ + \Gamma^2_+},
\eeq
where 
\bea
\nn \beta_e &=& 2 e^2 Q_q Q_\ell, \\ 
\nn \Gamma^n_\pm &=& [(g^q_L)^n \pm (g^q_R)^n][(g^\ell_L)^n \pm (g^\ell_R)^n].
\eea

For $q$=$u$ \& $\ell$=$e$, one obtains $\afb (\rm SM) = 0.6043$, in agreement with the SM curve in Fig.~\ref{fig:deviations}. 
Strikingly, one obtains a very similar value for $q$=$d$: $\afb (\rm SM) = 0.6365$.
Thus, at the hadron level, where the $\afb$ is roughly a weighted average of the up and down quark contributions, one expects $\afb (\rm SM)$ at high $\mll$ to fall between these two closeby values.

I now describe the CS method assuming the transverse momenta of initial state partons can be neglected in comparison to the high longitudinal momenta generated by the LHC. 
First, every event is boosted along the beam axis until the {\em dilepton} centre-of-momentum frame is found.
The direction of this boost is then taken to be the provenance of the quark, owing to its predominantly valence nature.
In this frame, the angle between the (anti)quark and (anti)lepton, $\thcs$, is defined as 
\beq
\cos \thcs = \frac{p^{\rm tot}_z}{|p^{\rm tot}_z|} \cdot 2 \cdot  \frac{p^{\ell^-}_+ p^{\ell^+}_- - p^{\ell^+}_+p^{\ell^-}_-}{\mll^2}~,
\label{eq:cs} 
\eeq

where $p^i_\pm \equiv (E^i \pm p^i_z)/\sqrt{2}$ and $p^{\rm tot}_z$ is the total dilepton longitudinal momentum.
Events with $\cos \thcs > 0 (< 0)$ are tagged as forward (backward).
Re-writing the above quantities in terms of the $p_T$ and pseudorapidities of leptons $\eta^\pm$, one finds that the forward-backward asymmetry at a $p$-$p$ collider comes down to the charge asymmetry:
\beq
\afbexp = \frac{N(\Delta|\eta| > 0)- N(\Delta|\eta| < 0)}{N(\Delta|\eta| > 0)+ N(\Delta|\eta| < 0)},
\label{eq:afbcs} 
\eeq
where $\Delta|\eta| \equiv |\eta^-|-|\eta^+|$.

There is inevitable discrepancy between $\afbexp$ and the forward-backward asymmetry in the actual centre-of-momentum frame, hereafter denoted as $\afbpart$.
Since a fraction of (anti)quarks is always misidentified, some truly forward events are mistaken for backward, and vice-versa.
Thus, some forward-backward events are symmetrized, or put differently, $\afbexp$ is diluted with respect to $\afbpart$.
In the SM this could by a factor of $1.5 - 3$, depending on the $\mll$ and total dilepton rapidity.
The dilution factor is in general determined by the PDFs and the model.
Appendix~\ref{sec:csspec} provides an analytical calculation of this factor at a given $\mll$ for a given model, starting from the $\cos \thcs$ spectrum of hadron level cross-sections.

There are two other sources of uncertainty, as outlined in \cite{Grossi:2012hw}, although not as important as the one above.
I will re-address them in this section, showing that they can be neglected in our analysis.
These are:
(i) higher order QCD and QED phenomena such as ISR. 
Their effect is the migration of events across $\mll$ bins, which indirectly impacts $\afb (\mll)$.
Its influence is at its highest near the $Z$ peak and gets smaller with $\mll$.
This effect may be mitigated by the use of the Mustraal frame \cite{Richter-Was:2016mal,Richter-Was:2016avq}, or by integrating events into wide $\mll$ bins;
(ii) detector resolution, resulting in mismeasurement of $\mll$ and rapidities. 
The realistic $\afb$ measured after undergoing these effects is sometimes called ``uncorrected".

ATLAS and CMS have measured $\afbexp$ in $e^{+}e^{-}$ and $\mu^{+}\mu^{-}$ channels with $20~{\rm fb}^{-1}$ of data collected in the 8 TeV run of the LHC.
ATLAS presents these measurements for dilepton masses upto 4 TeV \cite{Aad:2014wca}, and
CMS for upto 2 TeV \cite{Khachatryan:2016yte}. 
CMS also divides events by their absolute rapidity 
\beq
y \equiv \half \log \left( \frac{E^{\rm tot}+p^{\rm tot}_z}{E^{\rm tot}-p^{\rm tot}_z} \right)~
\label{eq:y}
\eeq
into four bins: $[0, 1], [1, 1.25], [1.25, 1.5], [1.5, 2.4]$.
The backgrounds in these measurements are the same as in Sec.~\ref{sec:mll}, with $q\bar{q} \ra Z/\gamma^* \ra \ell^+\ell^-$ dominating.

To derive constraints on leptoquark models, I use the CMS measurements as they show data in more numerous bins. 
As in Sec.~\ref{sec:mll}, I generate $p p \ra \ell^+ \ell^-$ events in {\tt MadGraph5} with a common renormalization and factorization scale of $\mll$ using CTEQ6L1 PDFs.
I then bin them in $y$ as above.
For the LQ masses considered here, only high $\mll$ events (far from the $Z$ peak) are relevant.
Therefore, I use only the last $\mll$ bin in the CMS measurement, corresponding to $[500, 2000]$~GeV.
I then determine $\afbexp$ in the four $y$ bins using Eqs.~\ref{eq:cs} and \ref{eq:afbcs}.
I find that the SM $\afbexp$ so obtained matches the background provided by CMS \cite{Khachatryan:2016yte} to an excellent degree.
This, then, testifies that the secondary sources of the dilution of $\afb$ -- ISR and detector resolution -- are indeed negligible at high $\mll$.

The uncertainties are almost entirely statistical; only in the bin $y \in [1,1.25]$ in the $\mu^+ \mu^-$ channel, the systematics are somewhat relevant, albeit subdominant.
To obtain the signal statistical uncertainty,
I simply rescale the background statistical uncertainty 
 by the ratio $\sqrt{(N_{\rm SM}/N_{\rm NP})}$, where $N_i$ is the number of SM or New Physics events in the relevant bin.
With this information, I obtain a 95\% C.L. bound on the leptoquark parameters with Pearson's $\chi^2$ statistic (see \cite{Hewett:1997ce,Agashe:2014kda}) by computing
\bea
\nn \chi^2_{\rm NP} &=& \sum_{\rm bins} \frac{(\afb^{\rm obs} - \afb^{\rm NP})^2}{\delta_{\rm NP}^2}~, \\
\chi^2_{\rm SM} &=& \sum_{\rm bins} \frac{(\afb^{\rm obs} - \afb^{\rm SM})^2}{\delta_{\rm SM}^2}~,
\label{eq:chisqafb}
\eea
and seeking $\Delta \chi^2 = \chi^2_{\rm NP} - \chi^2_{\rm SM}$ = 5.99.
The results, plotted in Fig.~\ref{fig:limits} using green curves, will be discussed in Sec.~\ref{sec:results}.
In the next sub-section I discuss a quantity that complements $\afbexp$ as a description of $\ell^+\ell^-$ angular spectra and may be more suitable to identify New Physics signals.

\subsubsection{Centre-edge asymmetry}
\label{sec:otherangulars}

\begin{figure}
\begin{center}
\includegraphics[width=.45\textwidth]{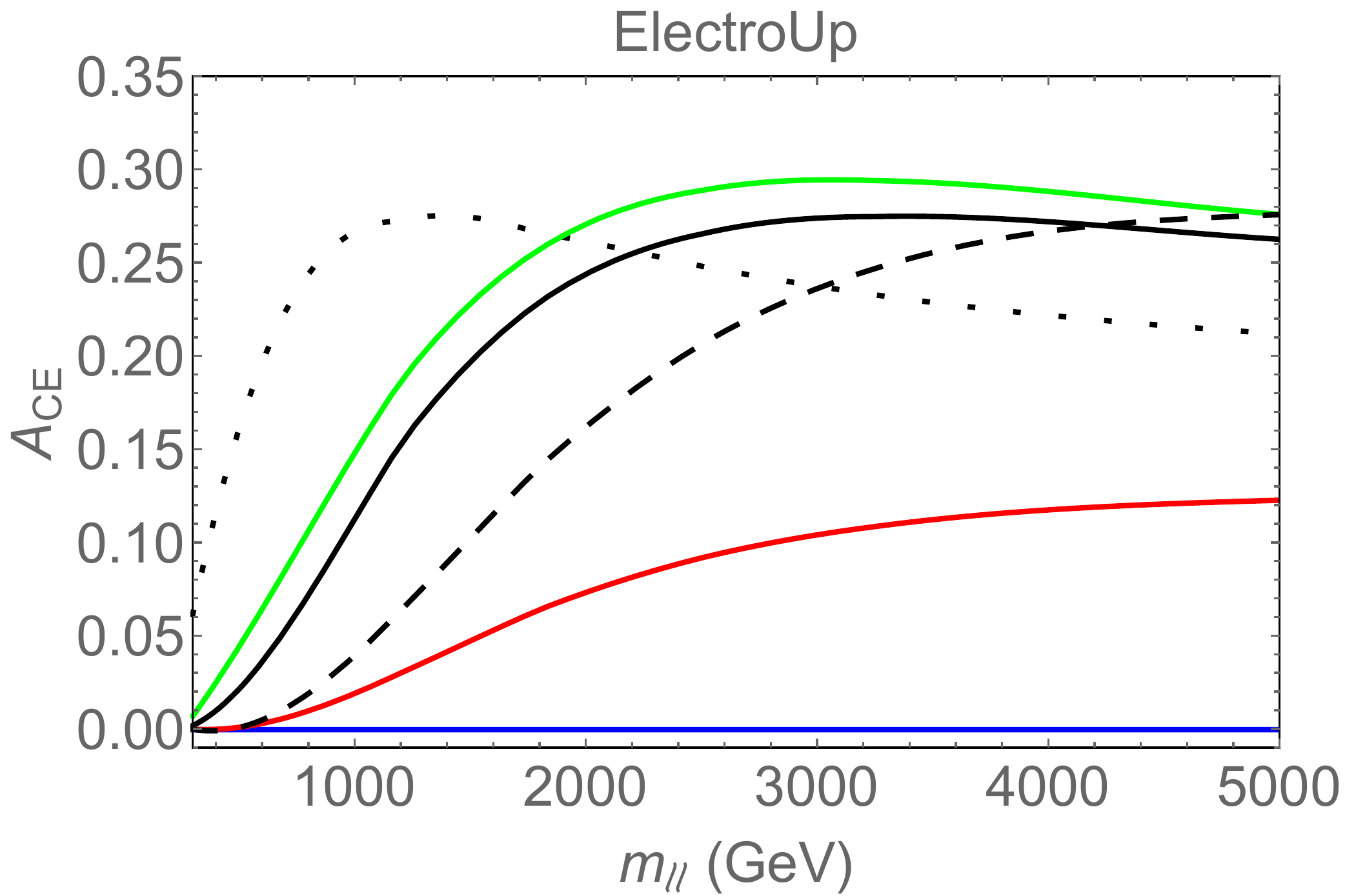} 
\caption{The centre-edge asymmetry $\ace$, as defined in Eq.~\ref{eq:ace}, with the colour code the same as in Fig.~\ref{fig:deviations}.
The $\ace$ is a boost invariant that can be chosen to vanish in the SM at some perturbative order, so that a non-zero value may easily signal new physics.
Here I have chosen $y_{\rm max} \ra \infty$ for simplicity and $y_0 = 3.854$ to set the SM value to zero. 
}
\label{fig:ace}
\end{center}
\end{figure}

\begin{figure*}
\begin{center}
\includegraphics[width=14cm]{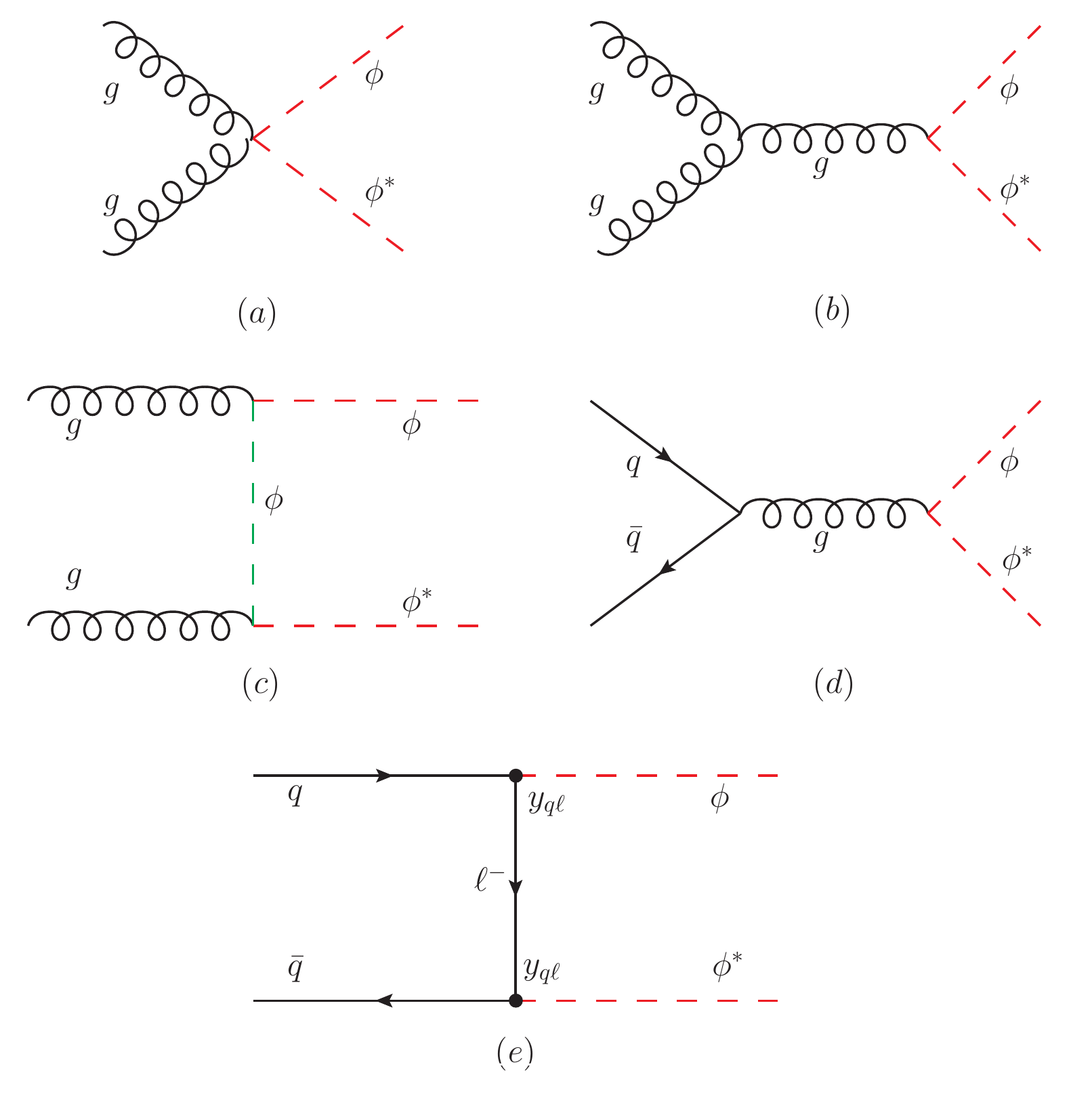} 
\caption{Feynman diagrams for pair production of leptoquarks, denoted by $\phi$.
Leptoquarks shaded red are produced on-shell, and those shaded green mediate production.
Diagrams (a)-(d) are QCD-driven while the diagram in (e) makes a Yukawa coupling-dependent contribution, subjecting the coupling to a mild constraint.
}
\label{fig:feynpair}
\end{center}
\end{figure*}

\begin{figure*}
\begin{center}
\includegraphics[width=15cm]{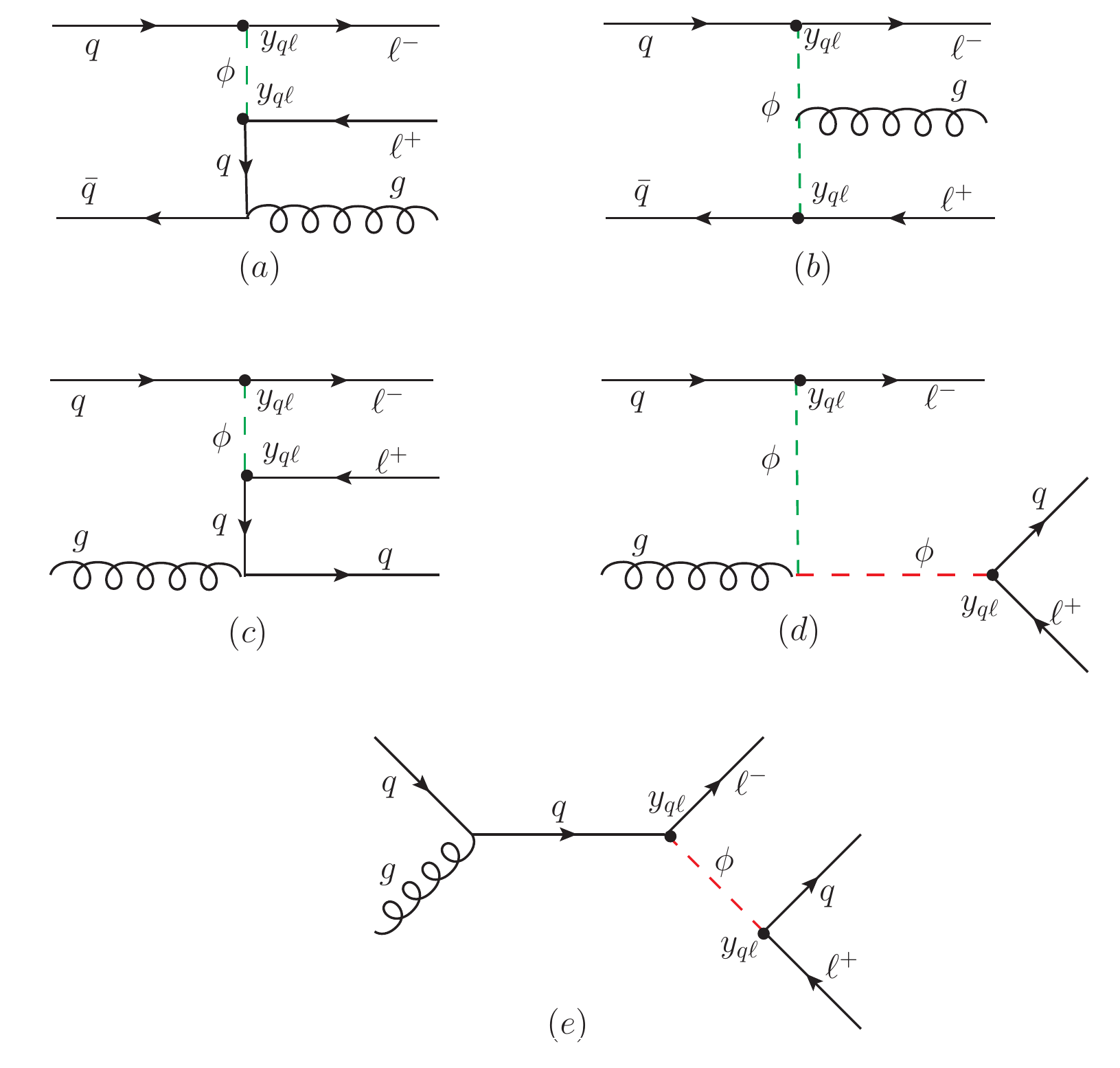} 
\caption{Feynman diagrams for single production of leptoquarks, denoted by $\phi$.
The LQs shaded red in diagrams (d) and (e) are produced on-shell; in these diagrams the LQ decay width in Eq.~\ref{eq:width} plays a major role in determining the signal cross-sections.
LQs shaded green mediate processes giving the $\ell^+\ell^-j$ final state.
Diagrams with all fermion arrows reversed are also present.
Unlike pair production, which is QCD-dominated and mostly unresponsive to LQ Yukawa couplings, this search channel is sensitive to both the couplings and masses of LQs. 
}
\label{fig:feynsingle}
\end{center}
\end{figure*}


The forward-backward asymmetry as a collider observable has its virtues.
In the angular spectrum of spin-1 $s$-channel mediation -- Eq.~\ref{eq:schangspec} -- one can determine that $a = 8\afb/3$. 
Thus the $\afb$ succinctly characterizes the (parabolic) $\cos\theta$ distribution by indicating the location of the minimum.
The SM $\afb$ also carries the telltale stamp of parity violation in the weak interaction. 
Indeed, the most accurate extractions of the weak mixing angle $s_W$ at high energies are made using $\afb$ measurements at the $Z$ pole, a topic to which I will return in Sec.~\ref{sec:discs}.

Yet the use of $\afb$ is perhaps outdated in the LHC era.
This may be argued with two reasons.

(i)  In $e^+ e^-$ colliders, as knowledge of the initial state directions fixes the sign of the beam axis,
the centre-of-momentum (CoM) frame scattering angle is always obtainable from final state pseudorapidities: $\cos\theta = \tanh(\Delta\eta/2)$, where $\Delta\eta = \eta^{\ell^-} - \eta^{\ell^+}$. 
From Eq.~\ref{eq:afbdef} one has
\beq
\afbpart = \frac{N(\Delta\eta > 0) - N(\Delta\eta < 0)}{N(\Delta\eta > 0) - N(\Delta\eta < 0)}~.
\label{eq:afblep}
\eeq
The benefit of all this is the basis on $\Delta \eta$, an invariant under longitudinal Lorentz boosts.
Thus, even if the lab frame differs from the CoM frame of the colliding beams, the phenomenologist is undeterred:
it is always straightforward to calculate quantities in the initial state CoM frame, and seldom obvious in other frames.

In contrast are the difficulties at a $p$-$p$ collider (and to an extent at a $p$-$\bar{p}$ collider) described in Sec.~\ref{sec:afb}. 
Crucially, $\afbexp$ as given in Eq.~\ref{eq:afbcs} is based on $\Delta|\eta|$, which is {\em not} invariant under boosts.
To make contact with data is then discouraging.
One must either derive (or be aware of) the not-so-evident equations in Appendix~\ref{sec:csspec}, or use a Monte Carlo phase space generator to work in the Collins-Soper frame.
The interested physicist is not always the intrepid physicist. 

(ii) The $\ell^+ \ell^-$ $\afb$ vanishes for the tree-level $\gamma$-mediated process (since QED respects parity) and is non-zero when the $Z$ interference is added.
This transition from zero to finite was an important attribute that helped us to indisputably glean the presence of ``new physics".
In the current era, the LHC probes energy scales far above the $Z$ mass, and in this regime the SM $\afb$ is more or less constant at $\simeq 0.6$.
Confoundingly, the ``uncorrected" version comes with an $\mll$-dependent dilution factor.
Although a treatment of backgrounds, data and errors could testify that the measured data agrees with the SM, one no longer enjoys the privilege of ``eyeballing" the data to immediately discern so.

In light of (i) and (ii), one may then ask: can we characterize LHC angular distributions with a concise quantity that is (a) frame-independent, and (b) vanishing in the SM?
Criterion (a) is effortlessly met by the simple use of variables that only contain $|\Delta \eta|$.
The absolute value eliminates the sign and hence uncertainties over initial state quark direction.
Analytical calculations in the centre-of-momentum frame are possible since $|\cos\theta| = \tanh(|\Delta\eta|/2)$.
Indeed, the LHC experiments already characterize jet angular distributions using the variable $\chi \equiv \exp (|\Delta \eta|)$.

Neither is criterion (b) difficult to fulfill.
In a $|\Delta \eta|$ distribution, one can always find a region containing exactly half the events.
The difference in population between this and the remaining region vanishes.
One quantity that potentially shows this feature is the centre-edge asymmetry, advocated in \cite{Diener:2009ee}.
It is defined as 
\bea
\nn \ace (\mll) &\equiv& \frac{\left[\int_0^{y_0} - \int_{y_0}^{y_{\rm max}}\right] d|\Delta\eta| (d^2\sigma/d|\Delta\eta|d\mll)}{\left[\int_0^{y_0} + \int_{y_0}^{y_{\rm max}}\right] d|\Delta\eta| (d^2\sigma/d|\Delta\eta|d\mll)} \\
\nn &=& \frac{N(0<|\Delta\eta|<y_0)-N(y_0<|\Delta\eta|<y_{\rm max})}{N(0<|\Delta\eta|<y_0)+N(y_0<|\Delta\eta| <y_{\rm max})}~. \\
\label{eq:ace}
\eea
The value of $y_{\rm max}$ is set by the size of the detector and the region chosen for analysis.
(E.g., $y_{\rm max} = 5$ for an analysis that demands lepton $|\eta| < 2.5$.)
One may then locate a $y_0$ such that the SM $\ace \ra 0$. 
Of course, one must compute this at a suitable perturbative order: we already know from studies of the $t$-$\bar{t}$ charge asymmetry \cite{Chatrchyan:2011hk,ATLAS:2012an} and dileptonic $\afb$ \cite{Baur:2001ze} that higher order corrections may slightly reshape angular spectra.
The measurement of a finite $\ace$ may then facilitate a prompt interpretation of new physics.
In addition, thanks to boost invariance, the values of $y_{\rm max}$ and $y_0$ can be mapped to corresponding $|\cos\theta|$ values in the CoM frame, enabling rapid analytic calculation.

I show the usefulness of this variable by plotting in Fig.~\ref{fig:ace} the LO $\ace$ produced by an {\tt ElectroUp} exchange as a function of $\mll$.
The colour code for this plot is the same as for Fig.~\ref{fig:deviations}.
Taking $y_{\rm max} \ra \infty$ for simplicity, the SM value here goes to zero for $y_0 = 3.854$, corresponding to $|\cos\theta| = 0.588$.
LQ exchange triggers a finite $\ace$, and careful measurements across $\mll$ may reveal the LQ mass and coupling strength.

The use of $\ace$ is revisited in Sec.~\ref{sec:discs}, where it will be shown that the spin of the LQ participating in DY production is instantly recognizable from the sign of $\ace$.





\subsubsection{$\cos\theta_{\rm CS}$ distributions}
\label{sec:costheta}

ATLAS has measured the $e^+e^-$ and $\mu^+\mu^-$ $\cos \thcs$ distributions at $\shatsq = 8$ TeV, $\lag = 20~{\rm fb}^{-1}$ \cite{Aad:2014wca}.
Lamentably, setting limits here is not as straightforward as the procedure used for $\mll$ spectra in Sec.~\ref{sec:mll}.
In the bins with $|\cos \thcs| > 0.6$, backgrounds involving $\gamma$-induced $\ell^+\ell^-$ production, $t$-$\bar{t}$, multijets and $W$+jets become comparable to the leading background ($Z/\gamma^*$-mediated DY).
The multijet and $W$+jets backgrounds are estimated by data-driven methods, and modelling them is beyond the scope of this work.
I will not attempt this analysis further.

\begin{figure*}
\begin{center}
\includegraphics[width=.45\textwidth]{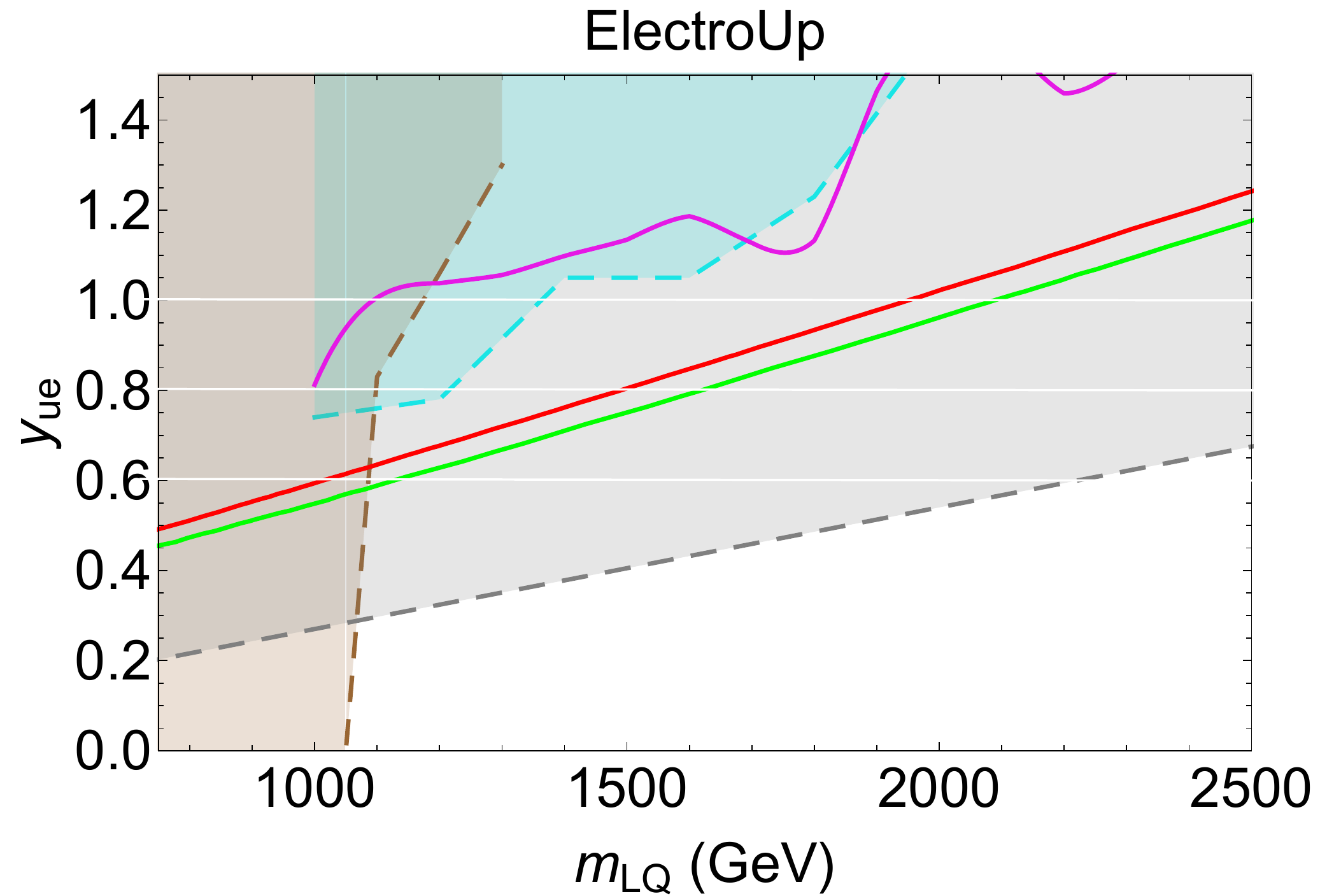} \quad
\includegraphics[width=.45\textwidth]{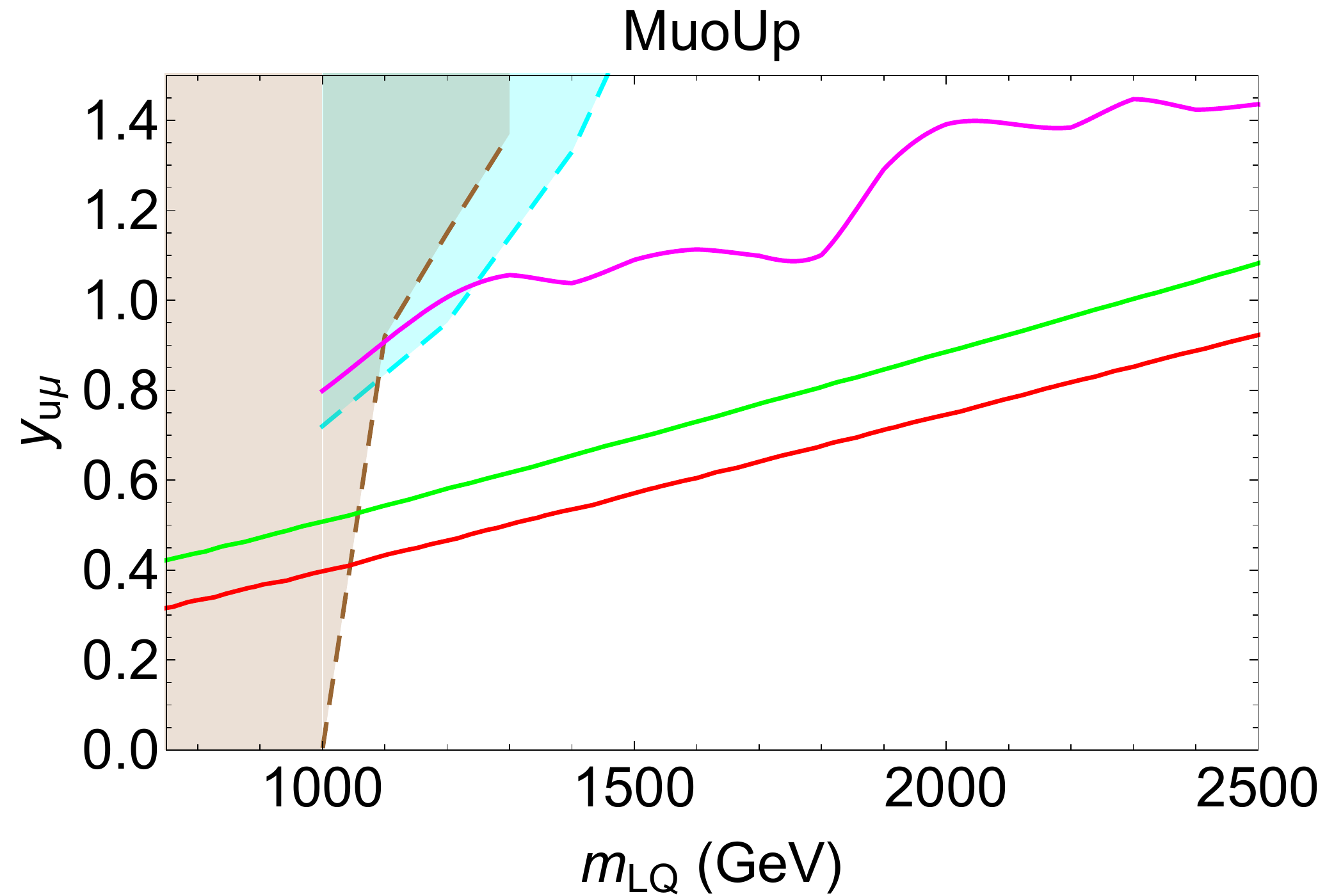}
\\
\vspace{1cm}

\includegraphics[width=.45\textwidth]{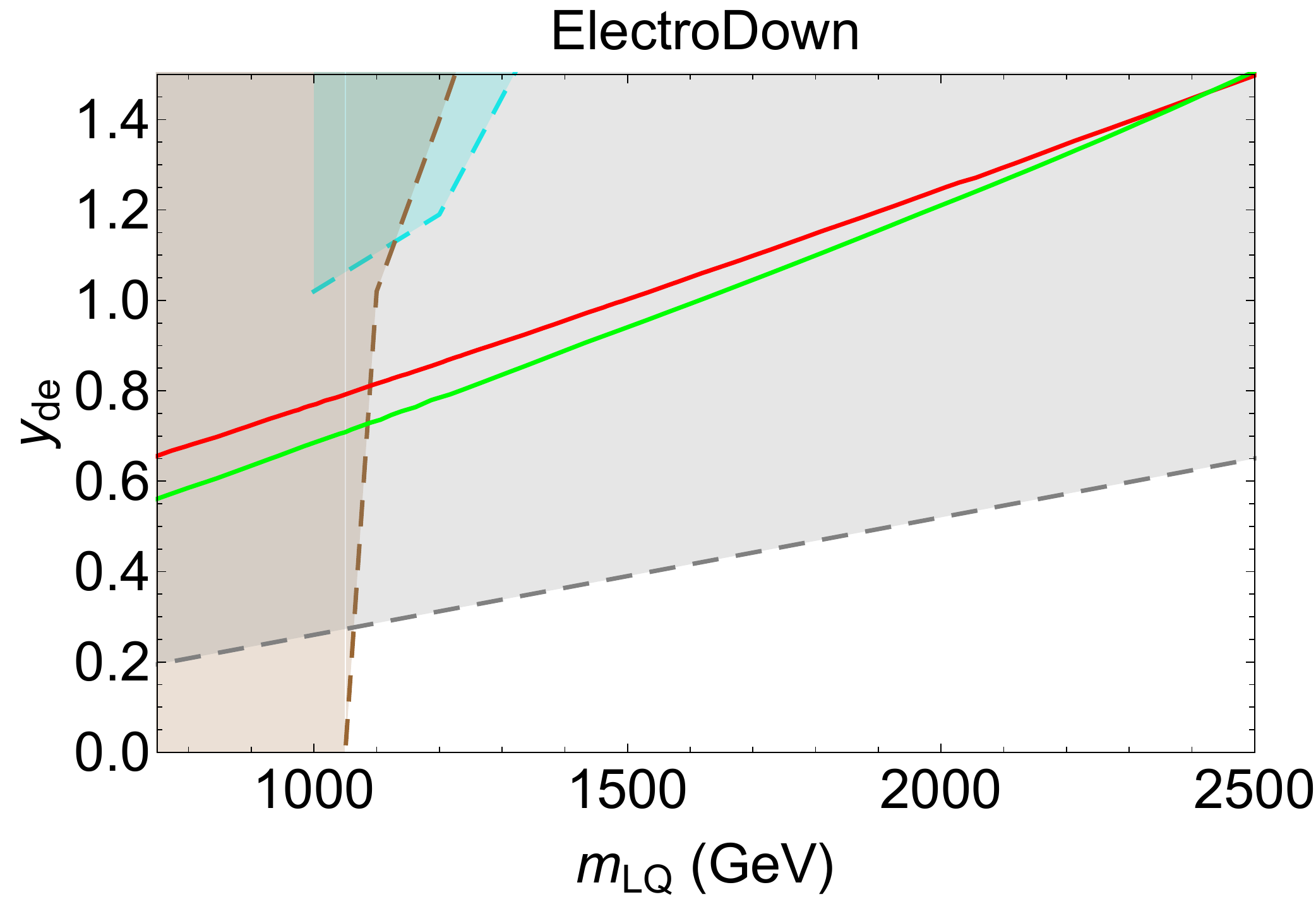} \quad
\includegraphics[width=.45\textwidth]{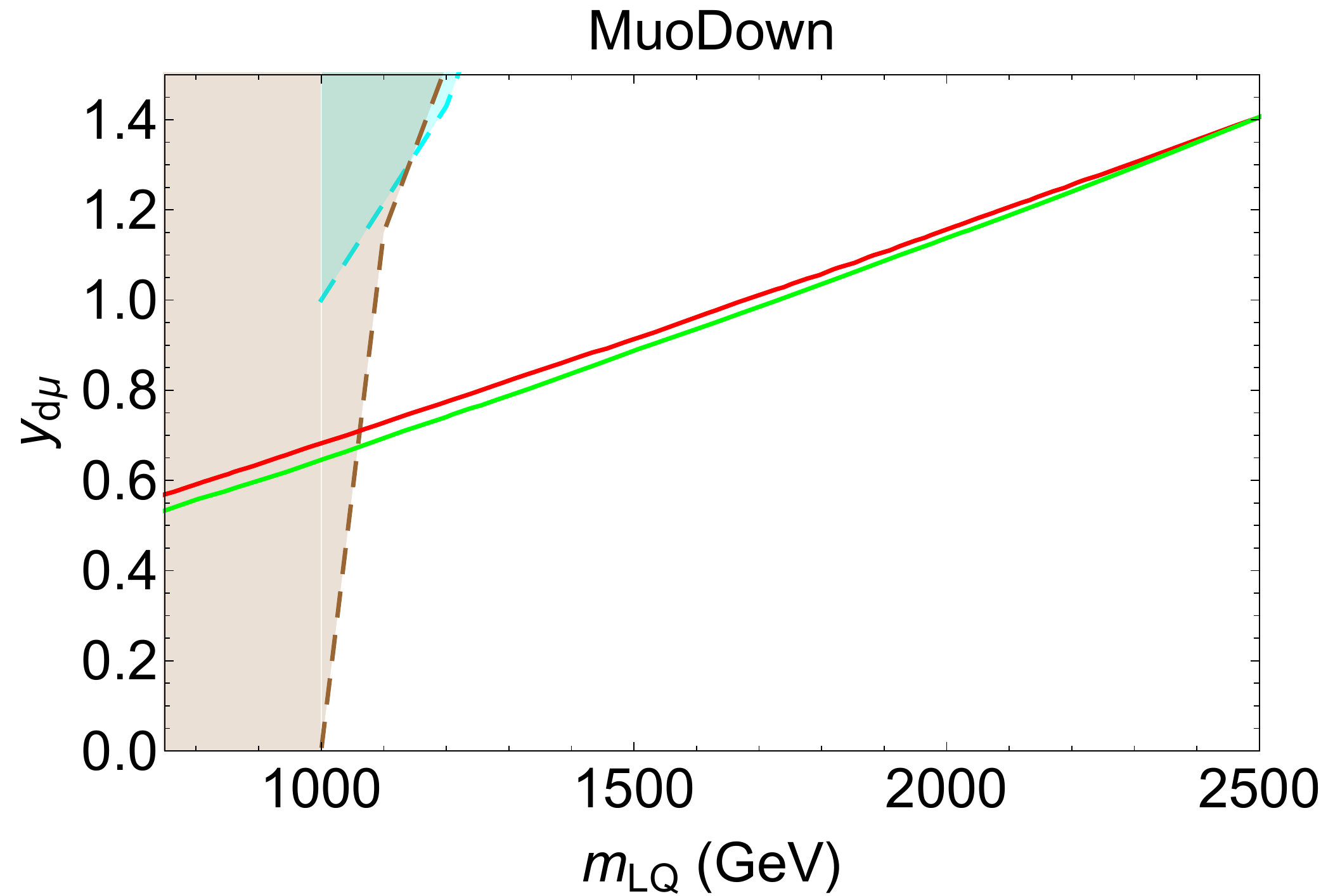}
\caption{Summary of all the constraints on LQ models considered in this work.
Regions shaded brown and cyan are excluded at 95\% C.L. by dedicated LHC searches in processes of LQ pair and single production respectively at $\sqrt{s} = 8$ TeV and $\lag = 20~{\rm fb}^{-1}$.
Regions shaded grey in the {\tt ElectroQuark} plots are excluded at $2\sigma$ by low-energy measurements of atomic parity violation.
The other curves are 95\% C.L. limits derived from LHC measurements of dilepton spectra at $\sqrt{s} = 8$ TeV, $\lag = 20~{\rm fb}^{-1}$.
The green curves are bounds from CMS measurements of the forward-backward asymmetry, and the red curves, from ATLAS data on $\mll$ distributions.
The magenta curves, provided for comparison, are limits extracted in Ref.~\cite{Wise:2014oea} from ATLAS bins of $\mll \geq 1.8$ TeV, where no events were observed. 
See text for further details.
}
\label{fig:limits}
\end{center}
\end{figure*}

\section{Other Leptoquark Probes} 
\label{sec:otherprobes}

Multiple experiments probe the leptoquark models in this work and can give interesting competition to the dilepton probes detailed in the previous section.
Sec.~\ref{sec:results} will show and discuss the corresponding limits.
 
Dedicated searches are ongoing at the LHC in processes of pair production (Fig.~\ref{fig:feynpair}) and single production (Fig.~\ref{fig:feynsingle}).
The first of these is dominated by QCD, with a small $\yql$-dependent contribution from the channel in Fig.~\ref{fig:feynpair}(e).
The second is completely sensitive to the $\yql$ couplings.
Precision measurements of atomic parity violation (APV) impose very stringent constraints on {\tt ElectroQuarks}, a fact poorly appreciated in leptoquark literature.
Finally, the measurement of $(g-2)_\mu$ may bear relevance to ${\tt MuoQuarks}$.
In Appendix~\ref{sec:aliterprobes}, I describe in detail the procedures I followed to map existing constraints onto my coupling-mass space.
I find that limits from direct production and APV experiments compete with the DY limits, while those from $(g-2)_\mu$ measurements do not apply in the relevant range of parameters.


\section{Results and Forecasts}
\label{sec:results}

I have assembled in Fig.~\ref{fig:limits} all the bounds arising from the probes discussed in Secs.~\ref{sec:probes} and \ref{sec:otherprobes}.
These bounds, where applicable, are shown for all four LQ models considered in this work, and plotted with a common key:

-- the red curves are 95\% C.L. limits from the ATLAS measurement of $\mll$ spectra at $\shatsq = 8$~TeV and $\lag = 20~{\rm fb}^{-1}$ \cite{Aad:2014cka}.
These limits were extracted in Sec.~\ref{sec:mll}. 

-- the magenta curves are the same, only as extracted in \cite{Wise:2014oea} using bins of $\mll \geq 1.8$ TeV, where zero events were observed.

-- the green curves are 95\% C.L. limits from the CMS measurement of $\afb$ at $\shatsq = 8$~TeV and $\lag = 20~{\rm fb}^{-1}$ \cite{Khachatryan:2016yte}.  
These limits were extracted in Sec.~\ref{sec:afb}.

-- the brown shaded regions are excluded at 95\% C.L. by pair production, as measured by ATLAS at $\shatsq = 8$~TeV and $\lag = 20~{\rm fb}^{-1}$ \cite{Aad:2015caa}. 
 These limits are extracted in Sec.~\ref{sec:pair} of Appendix~\ref{sec:aliterprobes}.

-- the cyan shaded regions are excluded at 95\% C.L. by single production, as measured by CMS at $\shatsq = 8$~TeV and $\lag = 20~{\rm fb}^{-1}$ \cite{Khachatryan:2015qda}. 
 These limits are extracted in Sec.~\ref{sec:single} of Appendix~\ref{sec:aliterprobes}.
 
 -- the grey shaded regions are excluded at 2$\sigma$ by the atomic parity violation measurement of Wood, et al \cite{Wood:1997zq}.
These limits are extracted in Sec.~\ref{sec:apv} of Appendix~\ref{sec:aliterprobes}.

I begin my discussion of these limits with the main results of this paper, describing where dilepton production stands in comparison to other probes:

(1) For $\mlq\gsim1$ TeV the measurement of both the $\mll$ distribution and $\afb$ constrain LQs better than pair production. 
This is because the cross-sections required to pair-produce LQs declines steeply with $\mlq$, whereas the modification to dilepton spectra from (even heavy) LQs can be appreciable, given a $\yql$ strong enough.
Below $\mlq = 1$ TeV, though, pair production, proceeding through QCD, restricts LQs better at small $\yql$, where the influence of LQs on dilepton production is faint.

(2) $\ell^+\ell^-$ measurements outdo single production.
This is because
 (a) the former involves two-body final states whereas the latter, three-body. 
Thus the cross-section of single production suffers a phase space cost,
(b) the final state $\ell^+ \ell^-$, pitted against $\ell^+ \ell^- j$, is much cleaner.
In different words, minimal theoretical and experimental uncertainties associate with purely leptonic final states, while the same cannot be said for single production -- the presence of the jet incurs larger uncertainties.

(3) In spite of the above virtues, dilepton production is as yet unable to compete with atomic parity violation in the {\tt ElectroQuark} models.
The great precision required to do so demands great collider luminosity, a topic to which I will return later in this section.

I now discuss every probe individually, beginning with dileptons. 

Several features are conspicuous in the red and green curves of Fig.~\ref{fig:limits}.  
Firstly, the red curves, representing the $\mll$ bounds derived here, run through smaller Yukawa couplings than the magenta curves, denoting the bounds derived by \cite{Wise:2014oea}.
The second limit is weaker because the authors of \cite{Wise:2014oea} obtained their limits using only data in bins with zero events. 

Secondly, the $\afb$ limits are (slightly) stronger than the $\mll$ spectrum limits in all but the {\tt MuoUp} models.
In the {\tt MuoUp} model, the $\mll$ distribution limit benefits from a down-fluctuation in the data, seen at 
$\mmm \simeq 750$ GeV.

Thirdly, the {\tt MuoQuark} models are better constrained than their {\tt ElectroQuark} counterparts.
As just mentioned, this can be put down in the case of the red curves to a down-fluctuation in dimuon production.
In the case of the green curves, the reason is the higher consistency of $\afb$ data with the SM in the $\mu^+\mu^-$ channel than in the $e^+e^-$ channel.

Lastly, the {\tt LeptoUp} models are better constrained than their {\tt LeptoDown} counterparts.
This is simply on account of the denser parton distributions of the up quark than the down in LHC protons.

Let me now summarize the dilepton constraints in mass ranges spared by LQ pair production.
Measurements of both the $\mll$ spectrum and the $\afb$ set upper bounds on LQ Yukawa couplings that increase monotonically with LQ mass.
At 95\% C.L., these bounds, for each LQ model in this work, are given by

\bea
\nn {{\tt ElectroUp}} & \\
\nn \mlq = 1.05 \ {\rm TeV}:  & \begin{cases} \yue \leq 0.63~[\mll], \\
  \yue \leq 0.58~[\afb]~. 
  \end{cases} \\
\nn  \mlq = 2.5 \ {\rm TeV}: & \begin{cases} \yue \leq 1.24~[\mll], \\
  \yue \leq 1.17~[\afb]~. 
  \end{cases} \\
  \nn  {{\tt ElectroDown}} & \\
\nn \mlq = 1.05 \ {\rm TeV}: & \begin{cases} \yde \leq 0.81~[\mll], \\
  \yde \leq 0.72~[\afb]~. 
  \end{cases} \\
\nn  \mlq = 2.5 \ {\rm TeV}: & \begin{cases} \yde \leq 1.5~[\mll], \\
  \yde \leq 1.5~[\afb]~. 
  \end{cases} \\
  \nn  {{\tt MuoUp}} & \\
\nn \mlq = 1.05 \ {\rm TeV}: & \begin{cases} \yum \leq 0.42~[\mll], \\
  \yum \leq 0.53~[\afb]~. 
  \end{cases} \\
\nn  \mlq = 2.5 \ {\rm TeV}: & \begin{cases} \yum \leq 0.92~[\mll], \\
  \yum \leq 1.08~[\afb]~. 
  \end{cases} \\
  \nn  {{\tt MuoDown}} & \\
\nn \mlq = 1.05 \ {\rm TeV}: & \begin{cases} \ydm \leq 0.72~[\mll], \\
  \ydm \leq 0.67~[\afb]~. 
  \end{cases} \\
\nn  \mlq = 2.5 \ {\rm TeV}: & \begin{cases} \ydm \leq 1.41~[\mll], \\
  \ydm \leq 1.41~[\afb]~. 
\end{cases}
\eea

Turning to pair production, one finds the excluded region very similar in all four LQ models.
This is because LQ pair production is dominantly QCD-driven in all four scenarios, with slight differences in the experimental analysis arising only from the reconstruction of the final state ($e^+e^-jj$ vs. $\mu^+\mu^- jj$).
In the {\tt ElectroQuark} ({\tt MuoQuark}) plots, the bound is $\mlq \leq 1050$ GeV ($\mlq \leq 1000$ GeV) near $\yql = 0$, and tends to higher masses as $\yql$ is increased -- a consequence of the increase in contribution of the production mode in Fig.~\ref{fig:feynpair} (e).
In the {\tt LeptoUp} models, the mass bound seems to saturate at 1300 GeV because ATLAS does not provide exclusion cross-sections beyond this point.
Through ampler parton densities of the up quark in the initial state proton, one finds the constraints in the {\tt LeptoUp} plots slightly stronger than the corresponding {\tt LeptoDown} plots. 

Constraints from single production are seen to reach heavier LQs than pair production.
This is because, compared to the on-shell production of {\em two} heavy LQs, the production of a single LQ costs less energy; higher LQ masses may be probed with the greater rates in which this results.
As the process is sensitive to the Yukawa coupling $\yql$, so is the mass reach -- the heavier the LQ, the stronger the $\yql$ required to compensate for the corresponding loss in cross-section (see Fig.~\ref{fig:feynsingle}).
This probe is seen to work best only in the {\tt ElectroUp} model: for $\yql = 1.5$, it excludes $\mlq \leq 1900$ GeV (c.f. $\mlq \leq 1300$ GeV excluded by pair production).
In the other three LQ models, the single production bound does not fall far from the pair production bound.
This is due in the {\tt MuoQuark} models to poor exclusion limits -- see Table B.2 in \cite{Khachatryan:2015qda} -- and in the {\tt ElectroDown} model to the smaller PDFs of the down quark.

The stiffest constraint on {\tt ElectroQuark}s is imposed by the APV measurement, seen to probe $\yql$-$\mlq$ space far more invasively than LHC single production.
Even for an LQ as heavy as 2.5 TeV, Yukawa couplings down to 0.6 are excluded.
The efficacy of this probe is a direct result of the precision achieved in low-energy experiments.
Due to difficulties in observing weak neutral currents in muonic atoms, parity violation has not been measured in these systems (yet). 
Were the task accomplished, it might emerge the best probe of {\tt MuoQuark} models \cite{Langacker:1990jf}.
I will discuss more on this theme in Sec.~\ref{sec:discs}.

\begin{figure*}
\begin{center}
\includegraphics[width=.45\textwidth]{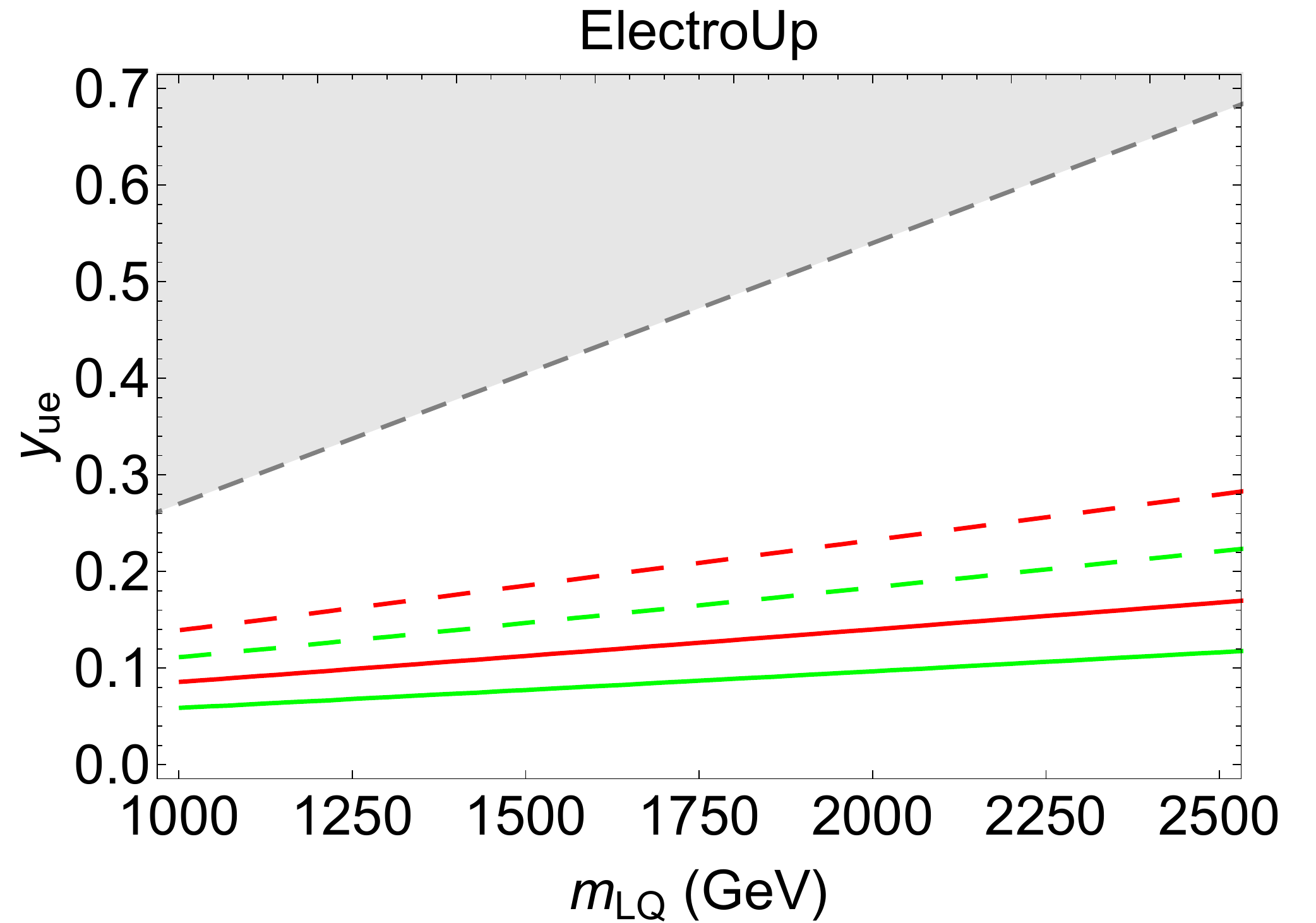} \quad
\includegraphics[width=.45\textwidth]{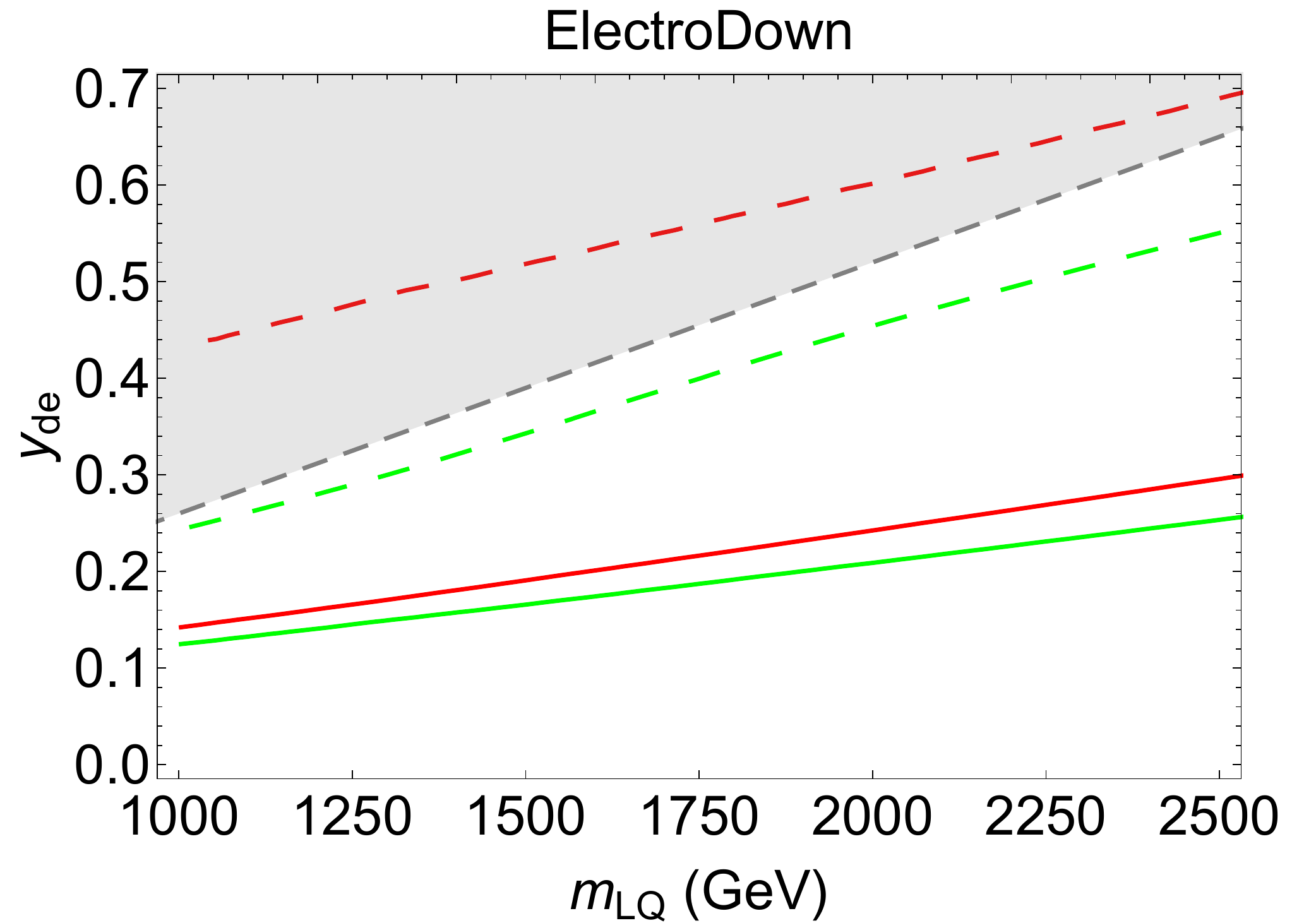}
\caption{95\% C.L. sensitivities of future LHC dileptonic measurements, compared with the current bound from 
atomic parity violation on {\tt ElectroQuarks} (shaded grey).
The green and red curves, as before, correspond to forward-backward asymmetries and $\mll$ spectra respectively.
The dashed (solid) curves correspond to an integrated luminosity of $300~{\rm fb}^{-1}$ ($3000~{\rm fb}^{-1}$).
These forecasts are made with pseudodata obtained by Poisson-fluctuating background estimates 100 times.
See text for further details.}
\label{fig:forecasts}
\end{center}
\end{figure*}

To conclude the discussion on constraints, I reiterate the main results of this work: for $\mlq \gsim 1$~TeV, dilepton distributions provide 
(i) stronger constraints than dedicated search strategies at the LHC, 
(ii) the strongest constraints to date on {\tt MuoQuarks}, while conceding to atomic parity violation in the case of {\tt ElectroQuarks}.

Can the dilepton channel ever overtake atomic parity violation? 
This is a question chiefly of collider luminosity
-- one now asks if the accumulated luminosities of future LHC runs can confer enough precision.
(As for the future of APV measurements, a number of experiments have been proposed, but their improvement over the previous precision of \cite{Wood:1997zq} is insignificant: see \cite{1111.4566} and the references in \cite{Agashe:2014kda}. 
I will therefore reuse constraints from \cite{Wood:1997zq}.)
Fig.~\ref{fig:forecasts} forecasts the 95\% C.L. sensitivity of the LHC for {\tt ElectroQuarks} at high luminosities.
The dashed (solid) curves correspond to $\lag = 300 \ {\rm fb}^{-1} \ (3000 \ {\rm fb}^{-1})$, and as before, the green (red) curves to sensitivities to the $\afb$ ($\mll$ spectrum) measurement.
To obtain these curves, I observed the following procedure, taking a leaf out of \cite{Altmannshofer:2014cla}'s book.
I first generated background events ($p p \ra Z/\gamma^* \ra e^+e^-$) at $\shatsq$ = 13 TeV using {\tt MadGraph5} over several bins of $\mll \geq$ 500 GeV.
I then generated ``data" with Poisson fluctuations around the background events in each bin.
I performed 100 of these pseudo-experiments.
Combing a grid of $\yql$ v. $\mlq$, I then generated signal events.  
Next, using Eqs.~\ref{eq:chisqmll} and \ref{eq:chisqafb} I found the 95\% C.L. sensitivity for all 100 sets of  pseudodata, assuming a systematic error of 6\%.
(The statistical error in $\afb$ was taken as $\sqrt{(1-\afb^2)/N}$.)
Finally, I averaged over the 100 $\Delta \chi^2$'s.
By this procedure, similar reaches will be obtained for {\tt Muoquarks}.

The results of this procedure can be summarized in the following 95\% C.L. sensitivities of either dileptonic probe. 
The upper bound in couplings is given for luminosities $(300 \ {\rm fb}^{-1}, 3000 \ {\rm fb}^{-1})$.
\bea
\nn {{\tt LeptoUp}} & \\
\nn \mlq = 1 \ {\rm TeV}:  & \begin{cases} \yql \leq (0.14,0.08)~[\mll], \\
  \yql \leq (0.11,0.06)~[\afb]~. 
  \end{cases} \\
 \nn \mlq = 2.5 \ {\rm TeV}: & \begin{cases} \yql \leq (0.28,0.17)~[\mll], \\
  \yql \leq (0.22,0.12)~[\afb]~. 
  \end{cases} \\
  \nn  {{\tt LeptoDown}} & \\
\nn \mlq = 1 \ {\rm TeV}: & \begin{cases} \yql \leq (0.42,0.14)~[\mll], \\
  \yql \leq (0.24,0.12)~[\afb]~. 
  \end{cases} \\
\nn  \mlq = 2.5 \ {\rm TeV}: & \begin{cases} \yql \leq (0.69,0.30)~[\mll], \\
  \yql \leq (0.56,0.26)~[\afb]~. 
  \end{cases} 
 \eea


These results show that high-luminosity LHC dilepton production can clearly probe {\tt ElectroQuarks} better than APV measurements, the only exception appearing in the $\mll$ spectrum at $300 \ {\rm fb}^{-1}$ for an {\tt ElectroDown}.
I have not attempted to project the LHC reach in single and pair production modes.
These rely sensitively on the cuts to be employed, and in any case one expects (on the strength of this paper's results) that the single production process will be a poorer probe of LQs than $\ell^+ \ell^-$ measurements. 
See \cite{Belyaev:2005ew} for a sensitivity study of combined single and pair production at $\shatsq$ = 14 TeV and $\lag = 300 \ {\rm fb}^{-1}$.

At the time of writing, the LHC collaborations have presented results of relevant searches at $\shatsq =$ 13 TeV. 
These include resonances in $\mll$ spectra (with $\lag \simeq 13\ {\rm fb}^{-1} $) \cite{ATLAS:2016cyf,CMS:2016abv} and LQs in pair production (with $\lag \simeq 3 \ {\rm fb}^{-1}$) \cite{Aaboud:2016qeg,CMS:2016imw}.  
There are no publications yet on $\afb$ measurements and single production searches at this energy and luminosity.
Since no meaningful comparisons can be made until we have these, I have not shown constraints using LHC Run 2 results.

\section{Summary and Scope}
\label{sec:discs}

\begin{figure}
\begin{center}
\includegraphics[width=.45\textwidth]{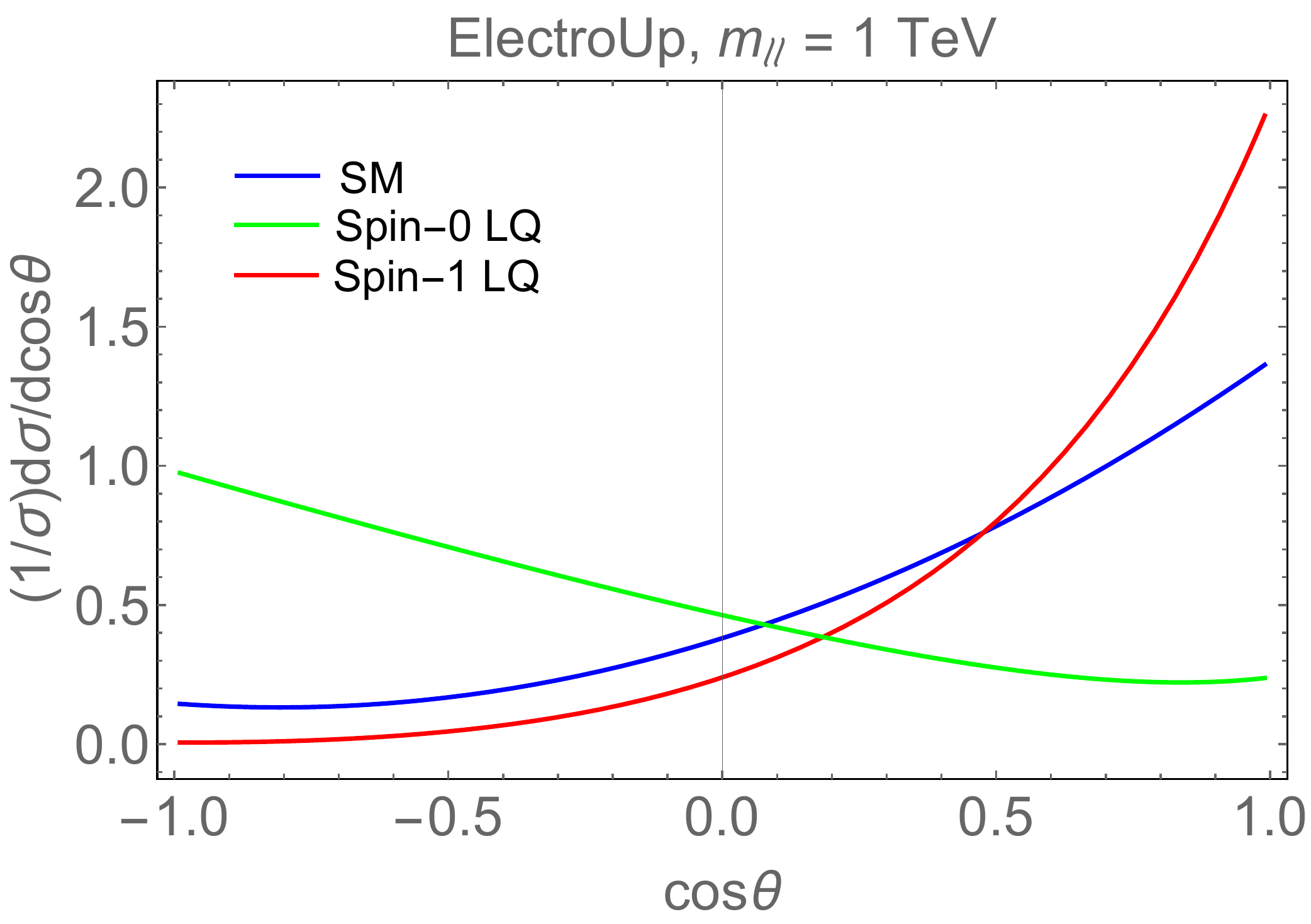}\\ \vspace{.4cm}
\includegraphics[width=.45\textwidth]{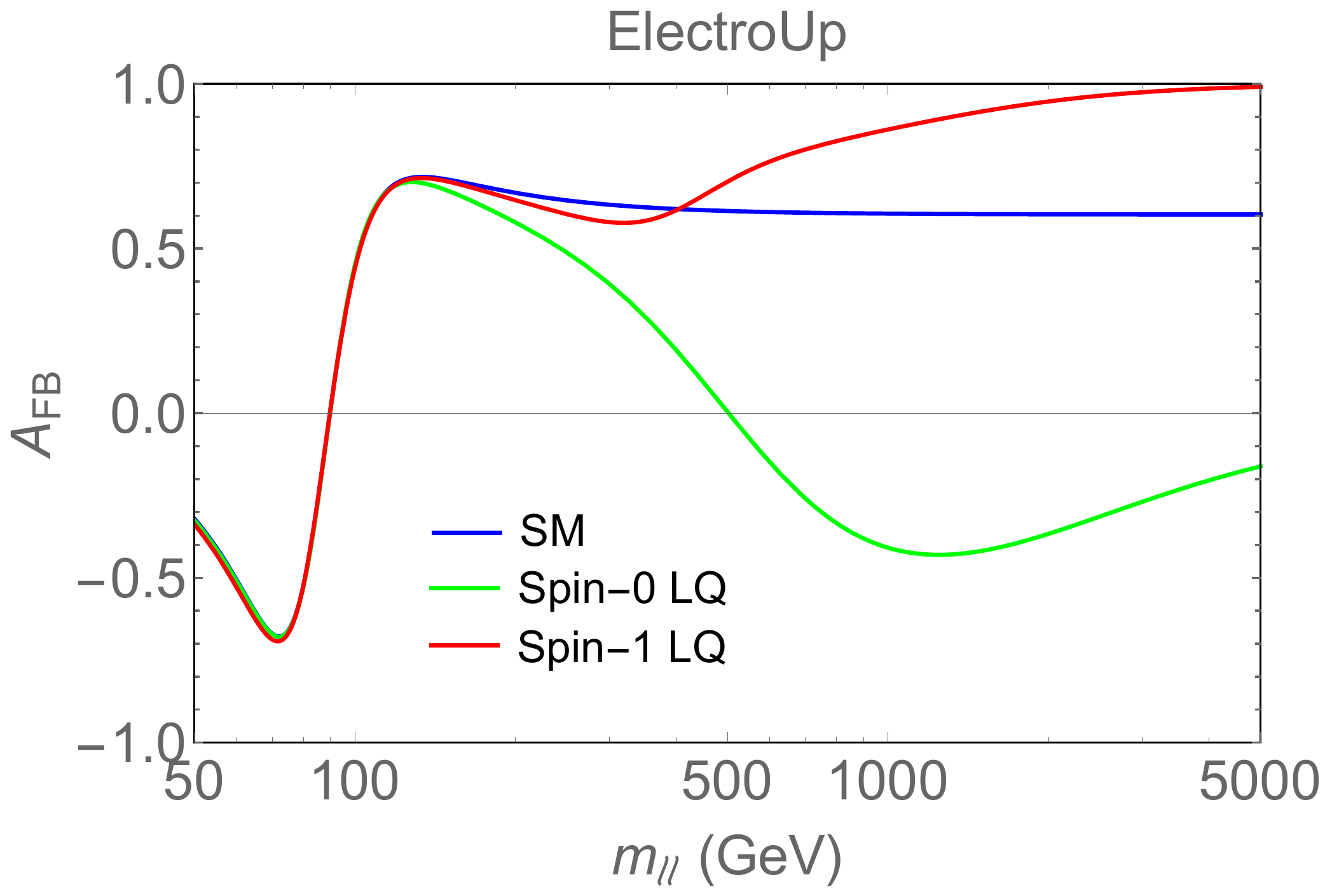}\\ \vspace{.4cm}
\includegraphics[width=.45\textwidth]{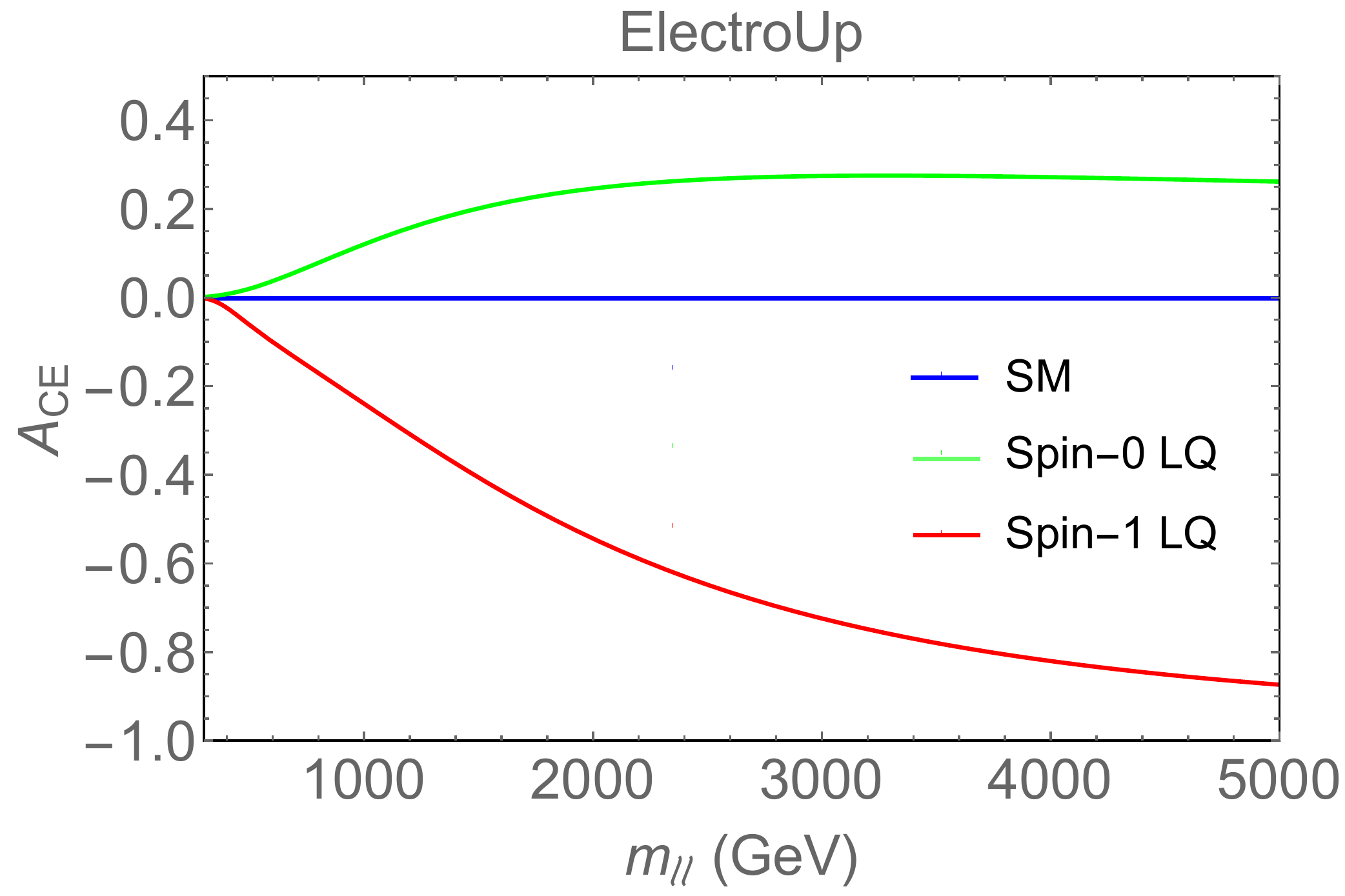}\\
\caption{$\ell^+\ell^-$ angular distributions can clearly mark the spin of LQs.
In these plots, blue: SM, green: scalar LQ and red: vector LQ.
The LQ coupling and mass are taken 1 and 1 TeV, for illustration.
The surest distinguisher is the centre-edge asymmetry $\ace$, carrying an opposite sign for either spin.
See Sec.~\ref{sec:discs} for more details.}
\label{fig:SvV}
\end{center}
\end{figure}

\begin{figure*}
\begin{center}
\includegraphics[width=15cm]{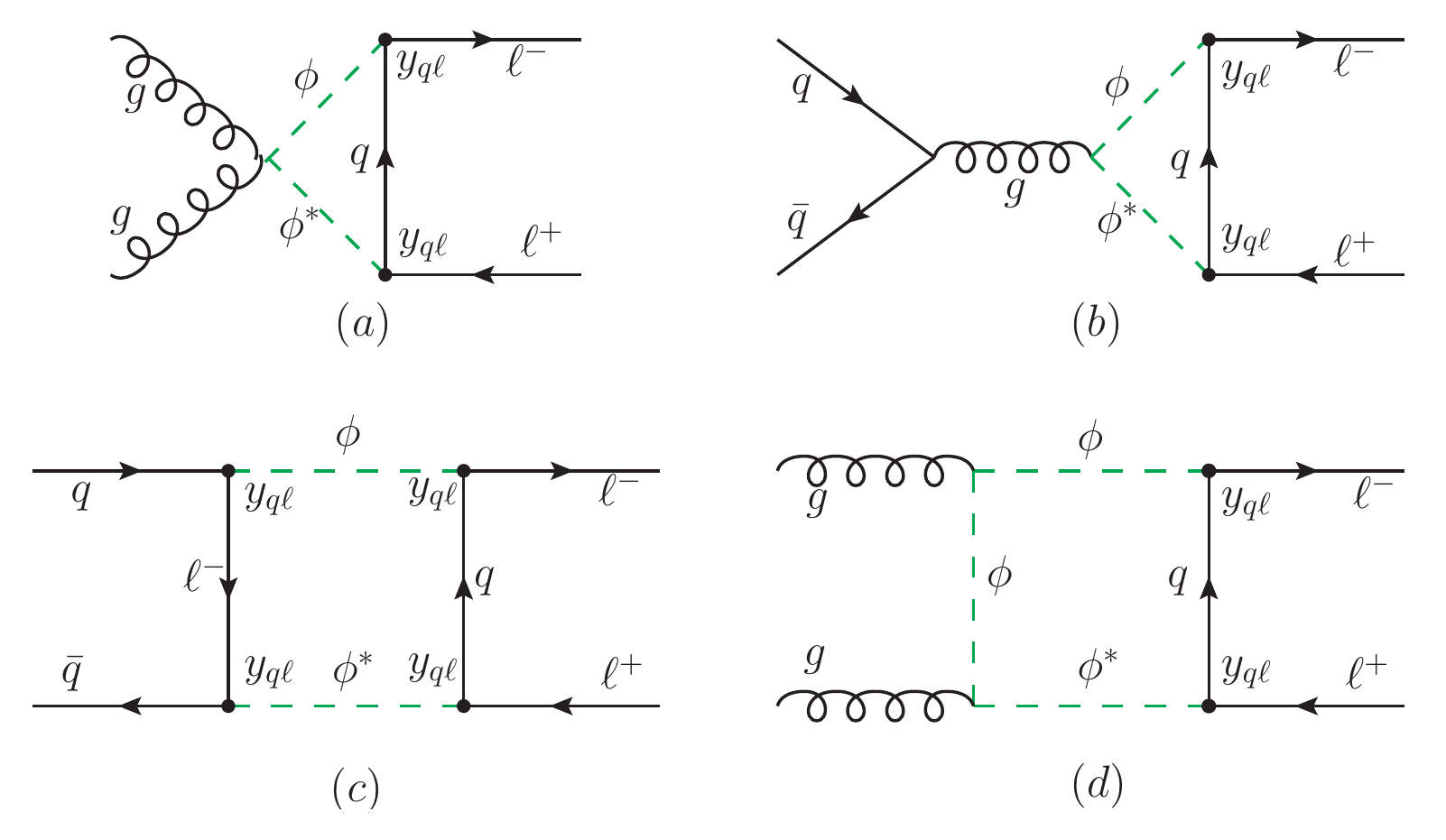} 
\caption{Feynman diagrams for radiative Drell-Yan processes that produce threshold effects, noticeable at large couplings.
See Sec.~\ref{sec:discs} for more details.}
\label{fig:feynloops}
\end{center}
\end{figure*}

In this work I showed how the celebrated virtues of dilepton production -- high rates, intelligible backgrounds, minimal theoretical and experimental uncertainties -- can be exploited to constrain non-resonant scenarios, specifically scalar leptoquarks exchanged in the $t$-channel of $q \bar{q} \ra \ell^+ \ell^-$.
The manner in which a $t$-channel scalar partitions dileptonic events in real space is quite unlike the Standard Model $s$-channel vectors. 
This difference can be picked up in collider measurements of angular distributions.
Thus the LHC measurements of dilepton angular spectra, found to agree with SM predictions, robustly constrain leptoquark parameters.

Indeed, I find the constraint from the $\ell^+ \ell^-$ forward-backward asymmetry typically tighter than that from the invariant mass spectrum.
These two constraints vie with, and often surpass, current limits on LQs from dedicated searches in processes of pair and single production.
When a leptoquark couples to the electron, low-energy measurements of atomic parity violation  provide better limits than LHC dilepton probes.
But future runs of the high-luminosity LHC can deliver enough precision in this channel to excel APV measurements.
Due to the use of different statistical tests to obtain limits from the $\mll$ spectrum and $\afb$, a statistical combination of the results was not attempted.

Also advocated in this paper is my view that the forward-backward asymmetry as a description of dilepton angular spectra is obsolete in the LHC era.
It is based on two facts: a meaningful $\afb$ at hadron colliders is only measurable in such special frames as the Collins-Soper frame, and it is non-zero at energies $>> M_Z$, where most searches for new physics are conducted.
In place of the $\afb$ is encouraged the use of the centre-edge asymmetry $\ace$, which is both frame-independent and vanishable in the SM, thus interfacing cleanly between collider measurement and theory calculation.

The implications of this paper's main results are many.
First, although a discovery of LQs in direct searches would be the clearest indication of their existence, it may be signalled first in the indirect probes of dilepton spectra.
If the LQs are so heavy as to not be pair-produced in sufficient numbers at the LHC, or couple to SM fermions so weakly as to not show up in single production processes, the sole signals may arrive in the Drell-Yan channel.
In such cases, information from both the kinematic and angular spectra may be required to reconstruct the LQ's mass, spin and couplings.

Second, the null results from LHC Run 1 for New Physics in the DY distributions of $e^+e^-$ and $\mu^+\mu^-$ channels marks out a region of parameters where dedicated LQ probes may not make a discovery in future searches.  
This statement is, of course, only true under certain circumstances.
The DY process is sensitive to the lepton-LQ-quark coupling $\yql$ (as is the single production process), while LQ pair production is QCD-dominated.
Therefore, the latter is obviously the leading probe when $\yql$ is small.  
Dilepton production is also a competitive probe only for LQs coupling the valence quarks $u$ and $d$ to electrons and muons.
To discover LQs of any other coupling structure, direct production (singly and in pairs) remains the best search strategy.
Due to this interplay of multiple probes, it is crucial to interpret LQ search data for all allowable coupling structures. 
As of now, the LHC experiments only present results on ``first", ``second" and ``third generation" LQs -- species that couple quarks with leptons of the same family number.
I would like to recommend the additional presentation of bounds on LQ species with cross-family couplings, such as the ones considered in \cite{Wise:2014oea} and in this paper.

Third, the conclusion that {\tt ElectroQuark} couplings are best bound by APV experiments while {\tt MuoQuark} couplings by LHC dimuon production underlines the importance of testing new sources of parity violation with precision experiments at low energies.
Of great interest would be to interrogate muonic systems at these scales and check if better limits than the LHC can be obtained.
The case for these measurements is compelling even outside the scope of this paper's results.
Intriguing discrepancies have been observed in $(g-2)_\mu$ \cite{Bennett:2006fi} and the proton charge radius obtained from the Lamb shift of muonic hydrogen \cite{Pohl:2010zza}.
While the observation of weak neutral currents in muonic atoms has remained infeasible, other avenues to probe parity violation have been proposed, such as the atomic radiative capture of muons \cite{McKeen:2012sh}.
If realized, these experiments may well become the frontier probe of parity-violating muon physics like {\tt MuoQuarks}.

While the focus of this paper was on the relevance of dilepton angular distributions to leptoquarks, these spectra are sensitive to other physics as well.
The weak mixing angle $s_W$ was extracted from LHC $\afb$ measurements at the $Z$ pole at $\sqrt{s} = 7$ TeV and found to be accurate to 0.44\% \cite{Agashe:2014kda}, with the error dominated by PDF systematics.
This uncertainty is already comparable to that from LEP measurements, 0.26\%, giving an indication of the precision achievable with $\afb$ at the LHC.
Measurements of ratios of cross-sections like $\afb$ and $\ace$ enjoy another advantage over $\mll$ spectra.
The lack of precise knowledge of the PDFs at high $x$ may limit a bump search, but may not pose as great a problem for $\afb$ or $\ace$, where the PDF uncertainty cancels to an extent in the ratio \cite{Accomando:2015cfa,1511.04573}.
These features may help these measurements become more sensitive to New Physics than $\mll$ distributions in future LHC runs.
Ref.~\cite{Accomando:2015cfa} makes use of these features to find that at LHC luminosities $\geq 30 \ {\rm fb}^{-1}$, $\afbexp$ may precede the $\mll$ spectrum as the discovery ground of a $Z'$ boson if the resonance is broad, and can give comparable sensitivities if its width is narrow.
Dilepton angles may have implications for the properties of dark matter as well.
One may inspect the sensitivity of LHC $\ell^+ \ell^-$ angular observables to the couplings and masses of dark matter (and its mediators) if they contribute to $q\bar{q} \ra \ell^+\ell^-$ through radiative processes, such as studied in \cite{Altmannshofer:2014cla}. 
Equally interesting to know would be the ability of these observables to discern DM spin and the chirality of DM's couplings to SM matter.
These possibilities are being explored in forthcoming work \cite{RCmonocline}.

Another exciting prospect is to find or corner supersymmetry in dilepton production distributions. 
This is possible in supersymmetric models with $R$ parity violation (RPV) \cite{Choudhury:2002av}. 
The RPV operator $\lambda'_{ijk} L_i Q_i D^c_k$ provides a leptoquark in the form of the down-type squark.
(In the usual LQ notation, this is the species $S_1$.)
The immediate constraints here are proton decay and unrestrained FCNCs that generally arise in RPV scenarios.
Bounds from proton decay may be mitigated if the baryonic RPV operator $\lambda''_{ijk}U_iD_jD_k$ and the respective soft term are suppressed,
while constraints from flavour violation depend on the pattern of $R$-parity breaking, the possibilities of which are reviewed in \cite{Barbier:2004ez}.
Another constraint immediately relevant in leptonic RPV comes from Higgs-slepton mixing.

The findings of this paper can be trivially extended to vector LQs.
Moreover, affairs may become doubly interesting if LHC dilepton spectra were to exhibit anomalies,
for not only can their leptoquark origins be tested, but also spin-0 LQ exchange can be untangled from spin-1. 
As already outlined in Eqs.~\ref{eq:tchangspecV}-\ref{eq:tchangspec}, we may perform this by studying angular distributions and asymmetries.
Fig.~\ref{fig:SvV} illustrates this clearly.
In these plots the blue curves correspond to the SM, and the green and red curves to scalar and vector LQs respectively, with unit coupling and LQ mass = 1 TeV.
For the vector LQ I choose the species $\widetilde{U}({\bf 3, 1},5/3)$, which mimics a {\tt LeptoUp}.
The top plot shows the angular distribution at $\mll = 1$ TeV, where qualitative differences between the two LQ spins are already apparent.
The middle plot tracks the partonic level $\afb$ as a function of $\mll$. 
The scalar LQ $\afb$ is always smaller than the SM value and even becomes negative, while the vector LQ $\afb$ by and large exceeds the SM value.
The difference is most manifest in the centre-edge asymmetry $\ace$ (defined in Eq.~\ref{eq:ace}), computed at partonic level and plotted as a function of $\mll$ at the bottom.
Here $y_0$ is chosen as 3.854 to make the SM value zero.
The $\ace$ unambiguously signals the spin by picking opposite signs for scalar and vector LQs.
 

As discussed in Sec.~\ref{sec:results}, the constraints on LQ couplings weaken monotonically with mass.
If LQs remain unseen to masses much higher than what was here considered, couplings as large as $\simeq$ 2 may be allowed.
If so, threshold effects due to loop-induced Feynman diagrams, such as those sketched in Fig.~\ref{fig:feynloops}, can result in ``monocline"-like features in dilepton spectra.
Such signatures were studied in the context of dark matter in \cite{Altmannshofer:2014cla}.
Constraints on the couplings now come primarily from the sharp-rise feature of the monocline, which may break the 
monotonicity against $\mlq$.
Notice also that unlike the simplified dark matter model interacting with quarks and leptons, LQs couple to gluons as well, resulting in added contributions to the $p p \ra \ell^+ \ell^-$ rates from the diagrams in Figs.~\ref{fig:feynloops}(a) and (d).
I leave for the future the exploration of scenarios outlined in this paragraph and the last.

In conclusion,
I hope that the encouraging results of this paper
and the richness of New Physics possibilities in dilepton angles at the LHC
will elevate their consequence from a profiling ground to a key discovery probe,
and place them in the same limelight as kinematic spectra.




\section*{Acknowledgments}

I am indebted to Ilja Dorsner for patiently explaining to me aspects of leptoquarks,
to Alexander Belyaev and Yue Zhang for their clarifying correspondence, 
to Carlos Alvarado, 
Joe Bramante,
Rodolfo Capdevilla, 
Antonio Delgado, 
Fatemeh Elahi, 
Paddy Fox, 
Roni Harnik, 
Graham Kribs,
Adam Martin and
James Unwin
for conversation and encouragement,
to Shyamala Venkataramani for suggesting that I write to CMS,
and to the public libraries of Hillsboro, OR, where this work was initiated and its better part completed, for their facilities.
Particular gratitude is owed 
to A. Delgado and A. Martin for reading the manuscript, 
to R. Capdevilla for checking a few calculations,
and to G. Kribs for raising the anomalous $g-2$ of the muon.
This work was supported by the National
Science Foundation under Grant 
No. PHY-1417118.

\appendix
\section{Angular spectrum in the Collins-Soper frame}
\label{sec:csspec}
Neglecting the transverse momenta of initial state partons, one may derive the angular distribution of events in the Collins-Soper reference frame (Sec.~\ref{sec:angspec}) using partonic cross-sections and the PDFs.
At hand are then spectrum-marking variables like $\afbexp$ for any model.

First, one re-writes Eq.~\ref{eq:cs} with (anti)lepton transverse momenta and pseudo-rapidities to get
\bea
\nn \cos \thcs &=& {\rm sgn}(p^{\rm tot}_z) \tanh (\Delta\eta/2) \\ 
 &=& {\rm sgn}(p^{\rm tot}_z) \ {\rm sgn}(p_q) \cos \theta,
 \label{eq:cscmlink}
\eea
obtaining a relation between the scattering angles in the CS and centre-of-momentum frames.
This equation clarifies that the positive beam axis -- the direction of the initial quark -- is chosen along the $p^{\rm tot}_z$ direction.
The PDFs determine
how probable the truth of this choice is.

Let me now denote the partonic differential cross-section for a quark flavour $q$ in the centre-of-momentum frame by 
\begin{equation*}
D_q(\ct) \equiv \frac{d\sigma_{q\bar{q}}}{d\ct} 
\end{equation*}
and, with $\tau = \mll^2/s$, define the partial luminosity functions
\bea
\nn L_q &\equiv& 2 \frac{\mll}{s} \left( \int_{\sqrt{\tau}}^1 \frac{dx}{x} f_q (x) f_{\bar{q}} (\tau/x) + \int_\tau^{\sqrt{\tau}} \frac{dx}{x} f_{\bar{q}} (x) f_q (\tau/x)  \right), \\
\nn L_{\bar q} &\equiv& 2\frac{\mll}{s} \left(  \int_{\sqrt{\tau}}^1 \frac{dx}{x} f_{\bar{q}} (x) f_{q} (\tau/x) + \int_\tau^{\sqrt{\tau}} \frac{dx}{x} f_q (x) f_{\bar{q}} (\tau/x) \right), \\
\label{eq:partlumfunx}
\eea
where $s$ is the collider energy and $f_i$ is the PDF of the parton $i$. 
The first (second) term captures events with greater momentum in the initial quark (antiquark).

At the hadronic level Eqs.~\ref{eq:cscmlink} and \ref{eq:partlumfunx} combine to give 
\beq
\frac{d^2\sigma_{pp}}{d\cos\thcs d\mll} = \sum_q (L_q D_q(\ct) + L_{\bar{q}} D_q(-\ct))~.
\label{eq:csspectrum}
\eeq

For plots of this spectrum in the SM and with other possible $s$-channel mediators, see \cite{Chiang:2011kq}.
From Eqs.~\ref{eq:afbdef} and \ref{eq:csspectrum} the CS forward-backward asymmetry is now constructed as
\beq
\afbexp (\mll) = \frac{\sum_q \left([\int_0^1 - \int_{-1}^0] d\ct D_q (\ct) [L_q - L_{\bar{q}}]\right)}{\sum_q \left([\int_0^1 + \int_{-1}^0] d\ct D_q (\ct) [L_q + L_{\bar{q}}]\right)}~,
\label{eq:afbtycal}
\eeq
where I have used 
\bea
\nn \int_{-1}^0 d\ct D_q(-\ct) &=&  \int_{0}^1 d\ct D_q(\ct),\\
 \nn \int_{0}^1 d\ct D_q(-\ct) &=&  \int_{-1}^0 d\ct D_q(\ct).
\eea

Comparing Eq.~\ref{eq:afbtycal} against its equivalent in the centre-of-momentum frame,
\begin{equation*}
\afb^{\rm CM} (\mll) = \frac{\sum_q \left([\int_0^1 - \int_{-1}^0] d\ct D_q (\ct) [L_q + L_{\bar{q}}]\right)}{\sum_q \left([\int_0^1 + \int_{-1}^0] d\ct D_q (\ct) [L_q + L_{\bar{q}}]\right)},
\end{equation*}
one discerns the dilution in $\afbexp$ coming from the PDFs.
As $L_{\bar{q}}$ diminishes with $\mll$, so does the dilution.

\section{Recasting Other Leptoquark Probes} 
\label{sec:aliterprobes}

This section explains the physics behind conventional probes of LQs
and describes in detail the procedures I followed to recast their measurements.


\subsection{Pair production}
\label{sec:pair}

Scalar LQs may be pair-produced at the LHC through channels depicted by the Feynman diagrams in Fig.~\ref{fig:feynpair}, each decaying subsequently to a lepton and a jet and producing such final states as $\ell^+\ell^-jj, \nu\bar{\nu}jj$ and $\nu \ell^\pm j j$.
The last two give rise to signals based on missing transverse energy and generally yield weaker bounds than the $\ell^+\ell^-jj$ channel.
The signatures relevant to this work are $e^+e^-jj$ and $\mu^+\mu^-jj$; the chief backgrounds are $Z/\gamma^* +$ jets, top pair production followed by leptonic decay, and diboson production.
Exclusion limits at 95\% C.L. on the signal production cross-section at $\shatsq = 8$ TeV are provided by CMS upto $\mlq =$ 1200 GeV \cite{Khachatryan:2015vaa} and by ATLAS upto $\mlq = 1300$ GeV \cite{Aad:2015caa} using 20 ${\rm fb}^{-1}$ of data.
{\tt ElectroUp}s with masses $\mlq \leq 1010$ and $\mlq \leq 1050$ GeV  are excluded by CMS and ATLAS respectively.
No limits are provided for {\tt ElectroDown}s, {\tt MuoUp}s and {\tt MuoDown}s, however the $\mu^+ \mu^- j j$ channel is used to constrain ``second generation LQs", i.e., LQs mediating interactions between second generation quarks and leptons.
These are excluded for leptoquark masses $\mlq \leq 1080$ GeV by CMS and $\mlq \leq 1000$ GeV by ATLAS.  
Both experiments assume that a negligible Yukawa coupling $\yql$, and thus omit the $t$-channel mode in Fig.~\ref{fig:feynpair}(e).

Since ATLAS enables a longer reach in $\mlq$ (and since CMS' results are similar), I will use their results for recasting purposes. 
First, setting $\yql = 0$, I compute the $\mlq$-dependent LO pair production cross-sections in {\tt MadGraph5} using CTEQ6L1 PDFs and by setting the common renormalization and factorization scale to $\mlq$.
With this information one may obtain the NLO differential $K$-factor as a function of $\mlq$, by comparing with the NLO signal cross-section provided by ATLAS\footnote{One may also obtain this by comparing against the NLO cross-sections for the pair production of top squarks with the rest of the superpartners decoupled \cite{SUSYWorkingGroup} 
.}. 
Next, I turn on $\yql$ and include the $t$-channel diagram in Fig.~\ref{fig:feynpair}(e) as a production mode.
The NLO cross-section is computed by assuming the $K$-factors obtained in the previous step. 
I then find, for various $\mlq$'s, the size of $\yql$ that saturates the ATLAS exclusion cross-section. 
Here I assume that the signal acceptance is not modified by the inclusion of the $t$-channel mode.



\subsection{Single production}
\label{sec:single}

CMS has provided bounds on LQ couplings and masses using processes that give $\ell^+\ell^-j$ final states \cite{Khachatryan:2015qda}. 
Signatures involving missing energy were not considered.
Proceeding through the Feynman diagrams in Fig.~\ref{fig:feynsingle}, these processes are collectively called ``single production".
Not all these diagrams amount to production of LQs, though -- diagrams (a), (b) and (c) only provide a means for achieving the $\ell^+\ell^-j$ final state through LQ exchange.
In these diagrams the decay width of the leptoquark does not appear in the amplitude, which I signify by drawing the leptoquark lines in green.
In diagrams (d) and (e) the leptoquark is produced on-shell and its width plays an important part; I have here shaded the leptoquark red.

The basic event selection criteria in the CMS analysis were these:
\bea
\nn  & &  p_T ({j}) > 125 \ {\rm GeV},\ |\eta ({ j})| < 2.4, \\
\nn & & p_T ({\rm \ell}) > 45 \ {\rm GeV},\ \ \ |\eta ({\rm \ell})| < 2.1,\ M_{\ell \ell} > 110 \ {\rm GeV}, \\
 & & S_T \equiv p_T (j_1) + p_T (\ell_1) + p_T (\ell_2) > 250~{\rm GeV}.
 \label{eq:basicselexion}
\eea
Events are picked with at least two leptons and at least one jet satisfying the criteria in the first two lines of Eq.~\ref{eq:basicselexion}, where $M_{\ell \ell}$ is the invariant mass of the lepton pair with the highest $p_T$.
In the third line, $p_T (j_1)$ is the $p_T$ of the leading jet, and $p_T (\ell_1)$ are $p_T (\ell_2)$ are the $p_T$ of the leading and second-leading leptons respectively.

The backgrounds come from the production of $Z/\gamma^*/W$ + jets, $t$-$\bar{t}$, $t$ + $X$, diboson + jets, and QCD jets with some jets misidentified as leptons.
To discriminate the signal from these backgrounds, a ``resonant selection" criterion was imposed on the events in order to favour the $\ell^+\ell^-j$ production modes through on-shell LQs (Figs.~\ref{fig:feynsingle} (d) and (e)). 
This criterion is
\beq
M_{\ell j} > f \times \mlq, \ \ \ S_T > S_T^0 (\mlq),
\label{eq:resselexion}
\eeq
where $M_{\ell j}$ is the lepton-jet invariant mass.
The constant $f$ was chosen as 0.75 (0.67) for the $eej$ ($\mu\mu j$) channel.
The threshold values $S_T^0$ were optimized for various $\mlq$ in either channel, and are provided in Tables B.1 and B.2 of \cite{Khachatryan:2015qda}.
The observed limits on signal cross-sections (at 95\% C.L.) after applying the criteria of Eqs.~\ref{eq:basicselexion} and \ref{eq:resselexion} are then derived as a function of $\mlq$ and provided in these tables.

To recast this search, I followed the procedure in \cite{Khachatryan:2015qda} (with some differences that I will discuss shortly).
First I generated the process $p p \ra \ell^+ \ell^- j$ in {\tt MadGraph5} using CTEQ6L1 PDFs.
I chose the common renormalization and factorization scale as $\mlq$.
I set the decay width of the LQs (see \cite{Dorsner:2014axa}) as
\beq
\wlq = \frac{\yql^2}{16\pi^2} \mlq ~.
\label{eq:width}
\eeq
Then I applied the selection criteria of Eqs.~\ref{eq:basicselexion} and \ref{eq:resselexion} and obtained the signal cross-section.
For a given $\mlq$, I varied $\yql$ until the exclusion cross-section in Tables B.1 and B.2 of \cite{Khachatryan:2015qda} was matched.

In \cite{Khachatryan:2015qda}, CMS provides a coupling vs. mass bound on ``first generation" LQs, choosing an LQ species that couples an up quark with an electron. 
The careful reader will notice a 20-25\% discrepancy between this bound and that provided in Fig.~\ref{fig:limits} for an {\tt ElectroUp}: my bound is somewhat looser.
In light of this, it is worth one's while to peer closer at the differences 
in the analyses performed by CMS and I.
\begin{itemize}

\item The LQ species used in \cite{Khachatryan:2015qda} is $S_1$ (see \cite{Dorsner:2016wpm}) with electric charge 1/3.
This LQ violates $B$ and $L$.
The relevant interaction is $\yql C \bar{u} S_1 P_R \ell$.
The LQ species I use are $R_2$ and $\tilde{R}_2$ with electric charge 5/3 and 2/3 respectively, and they conserve $B$ and $L$. 
The relevant interactions are in Eqs.~\ref{eq:Lag1} and \ref{eq:Lag2}.    


\item The effect of the decay width (Eq.~\ref{eq:width}) is not mentioned in \cite{Khachatryan:2015qda}. 

\item $\ell^+ \ell^- j$ production modes involving the diagrams in Figs.~\ref{fig:feynsingle} (a) and (b) were not mentioned in \cite{Khachatryan:2015qda}.

\item A minor difference. 
The mention of renormalization and factorization scale used is omitted in \cite{Khachatryan:2015qda}. 
However, in private correspondence with CMS I was informed that this scale was set to $\mlq$.

\end{itemize}

These differences may adequately explain the discrepancy between CMS and this work.
A thorougher investigation is outside my scope.
I should like to emphasize here that even if I had presented the bounds provided by CMS, they would be weaker than the Drell-Yan limits in the region not excluded by pair production. 


\subsection{Atomic parity violation}
\label{sec:apv}

The 1997 measurement of parity non-conservation at low energies by Wood, et al \cite{Wood:1997zq} in cesium-133 with an experimental error of 0.35\% is the state of the art. 
From this measurement was extracted the nuclear weak charge \cite{Dzuba:2012kx}:
\bea
\nn Q_W^{\rm expt} ({\rm Cs}) &=& -72.58 \pm 0.43, \\
Q_W^{\rm SM} ({\rm Cs}) &=& -73.23 \pm 0.20,
\eea
which amounts to a 1.5$\sigma$ deviation from the SM.
The Wood, et al measurement is a stringent test of new sources of atomic parity violation (APV) beyond the SM.
One understands this from the following effective Lagrangian below the EW scale \cite{Dorsner:2014axa}:

\beq
\lag_{\rm APV} = \frac{G_F}{\sqrt{2}} \bar{e}\gamma^\mu\gamma^5 e [C_{1u} \bar{u}\gamma_\mu u + C_{1d} \bar{d} \gamma_\mu d]~.
\eeq

In terms of the Wilson co-efficients, atomic number $Z$ and number of neutrons $N$, the nuclear weak charge is given by 
\begin{equation}
Q_W(Z,N) = -2[(2Z+N)C_{1u} + (Z+2N)C_{1d}]~.
\label{eq:nucweakchar}
\end{equation}

The SM contributions to the co-efficients come from $Z$ exchange: $C^{\rm SM}_{1u} = -1/2 + 4/3 s_W^2$ and $C^{\rm SM}_{1d} = 1/2 - 2/3 s_W^2$.
If one assumes the co-efficients receive some contribution from new physics such that $C_{1q} = C^{\rm SM}_{1q}+ \delta C_{1q}$, then from Eq.~\ref{eq:nucweakchar} the new contribution to $Q_W$ is
\beq
\delta Q_W(Z,N) = -2[(2Z+N)\delta C_{1u} + (Z+2N) \delta C_{1d}]~.
\eeq

Now leptoquark propagation gives a contribution to the coefficients given by 
\beq
\delta C_{1(u/d)} = \frac{\sqrt{2}}{G_F} \frac{|y_{(u/d)e}|^2}{8\mlq^2}~.
\eeq
Hence departures from the measured $Q_W$ brought about by LQs can be translated to limits on LQ parameters.
It was found that requiring no more than a 2$\sigma$ deviation from the measured value gives \cite{Dorsner:2014axa} 
\bea
\nn |y_{ue}| &\leq&  0.27\left(\frac{\mlq}{1 \ {\rm TeV}}\right)~, \\
 |y_{de}| &\leq& 0.26 \left(\frac{\mlq}{1 \ {\rm TeV}}\right)~.
\eea


\subsection{Muon anomalous magnetic moment}
\label{subsec:g-2}

There is an enduring 3$\sigma$ discrepancy between the value of $a_\mu \equiv (g-2)_\mu/2$ measured by the E821 experiment at BNL and its SM prediction \cite{Bennett:2006fi}.
Loops involving LQs can modify the theoretical prediction; their contribution can be calculated using
the formulae in \cite{Dorsner:2016wpm}.  
When applied to {\tt MuoQuark}s, I find that the discrepancy is further widened.
For a {\tt MuoDown}, the contribution is negligible due to an accidental cancellation in the loop functions.
For a {\tt MuoUp}, demanding that the discrepancy be no larger than 5$\sigma$ yields constraints that lie well outside the range of parameters presented in this work.  
More precisely, requiring $\yum \leq 2$ gives $\mlq \geq 350$~GeV, a very weak constraint compared to the other bounds here.









\end{document}